\documentclass{aa}  

\usepackage{natbib}
\bibpunct{(}{)}{;}{a}{}{,} 
\usepackage{caption}
\usepackage{multirow}
\usepackage{array}
\usepackage{pdfpages}
\usepackage{graphicx,times}             

\usepackage{graphicx}
\usepackage{txfonts}
\begin{document}

   \title{Dielectric properties and stratigraphy of regolith in the lunar South Pole-Aitken basin: Observations from the Lunar Penetrating Radar}

   \author{Jianqing Feng
          \inst{1,\ 2},\
          Matthew. A. Siegler\inst{1,\ 2},\and
          Mackenzie N. White\inst{1,\ 2}
          }

   \institute{Planetary Science Institute, 1700 East Fort Lowell, Tucson, AZ 85719, USA\\
              \email{jfeng@psi.edu}
         \and
             Roy M. Huffington Department of Earth Sciences, Southern Methodist University, 6425 Boaz Lane, Dallas, TX 75205, USA\\
             }


 
  \abstract
   {}
   {We examine data obtained by the Lunar Penetrating Radar (LPR) onboard the Chang'E-4 (CE-4) mission to study the dielectric properties and stratigraphy of lunar regolith on the far side of the Moon. }
   {The data collected from January 2019 to September 2020 were processed to generate a 540 m radargram. The travel velocity of the radar signal and the permittivity of the regolith were deduced from hyperbolas in the radargram. As CE-4 LPR detected distinct planar reflectors, we evaluated the dielectric loss from the maximum penetration depth based on the radar equation. The derived dielectric properties are compared with the measurements of Apollo samples and Chang'E-2 microwave radiometer observations.}
   {The results suggest that regolith at the landing site has a permittivity of 2.64--3.85 and a loss tangent of 0.0032--0.0044, indicating that the local regolith is composed of a fine-grained, low-loss material that is much more homogeneous than that found at the Chang'E-3 landing site. The total thickness of weathered material is $\mathrm{\sim}$ 40 m, with several regolith layers and a buried crater identified in the reconstructed subsurface structure. }
   {These layers clearly record a series of impact events from the adjacent regions. We suggest that the top layer is  primarily made up of the ejecta from a large crater 140 km away. In contrast, the material source of other thinner layers comes from nearby smaller craters.}


   \maketitle
%
\section{Introduction}
Impact cratering has shaped the surface of the Moon throughout the Moon's evolutionary history. Due to continuous impact gardening, nearly the entire lunar surface consists of a layer of regolith that completely covers the underlying bedrock \citep{Heiken}. Generally, the regolith has an average thickness of 4--5 m in the maria and 10--15 m in the highlands \citep{Mckay}. Subsurface structures in the regolith record substantial evidence of impact cratering processes and volcanic eruptions. In 2014, Chang'E-3 (CE-3) mission surveyed the lunar subsurface at its landing site in northern Mare Imbrium using the Lunar Penetrating Radar (LPR) --- the first ground penetrating radar (GPR) onboard a rover on the Moon \citep[e.g.,][]{Su,Feng2017}. In 2019, Chang'E-4 (CE-4) landed on the eastern floor of Von Kármán crater on the far side of the moon. The Yutu-2 rover of CE-4 mission carried an LPR with the same configuration as CE-3 \citep[e.g.,][]{Li,Lai2019,Ding}. The LPR is an ultra-wideband carrier-free pulse radar with two channels. The first channel installed on the back of the rover operates at a center frequency of 60 MHz with a bandwidth of 40 MHz, while the second channel installed on the underside of the rover works at a center frequency of 500 MHz with a bandwidth of 500 MHz \citep{Zhanghongbo,Li}.     

 Studies have investigated the thickness of lunar regolith in various ways. From the Lunar Orbiter program in the 1960s to the Lunar Reconnaissance Orbiter (LRO) mission throughout the last 12 years, many researchers have used the morphology of small craters and crater counting techniques to constrain the regolith thickness \citep[e.g.,][]{Bart2011,Bart2014,Oberbeck1967,Oberbeck1968,Shoemaker,Wilcox}. During the Apollo 15, 16, and 17 missions, astronauts took in-situ measurements at the landing sites. The maximum penetration depth of the drills of Apollo 15, 16, and 17 are 2.4 m, 2.25 m, and 2.95 m, respectively \citep{Allton}. Although the drills did not touch the base rock, they provided an upper boundary constraint on the regolith thickness at the landing sites. Passive seismic experiments estimated the thickness from the shear-wave resonance at the Apollo 11, 12, and 15 sites to be 4.4, 3.7, and 4.4 m, respectively \citep{Nakamura}. As microwave and radio wave signals can penetrate the regolith much deeper than optical signals, orbital radar technology has been effectively applied to detect lunar subsurface structures since the 1970s \citep[e.g.,][]{Phillips,Spudis,Ono}. Earth-based radio telescopes are also used to observe the near side of the Moon and calculate the regolith thickness and associated dielectric properties \citep[e.g.,][]{Thompson,Shkuratov,Fa,compbell}.

   \begin{figure}[htb]
    \includegraphics[width=8cm,angle=0]{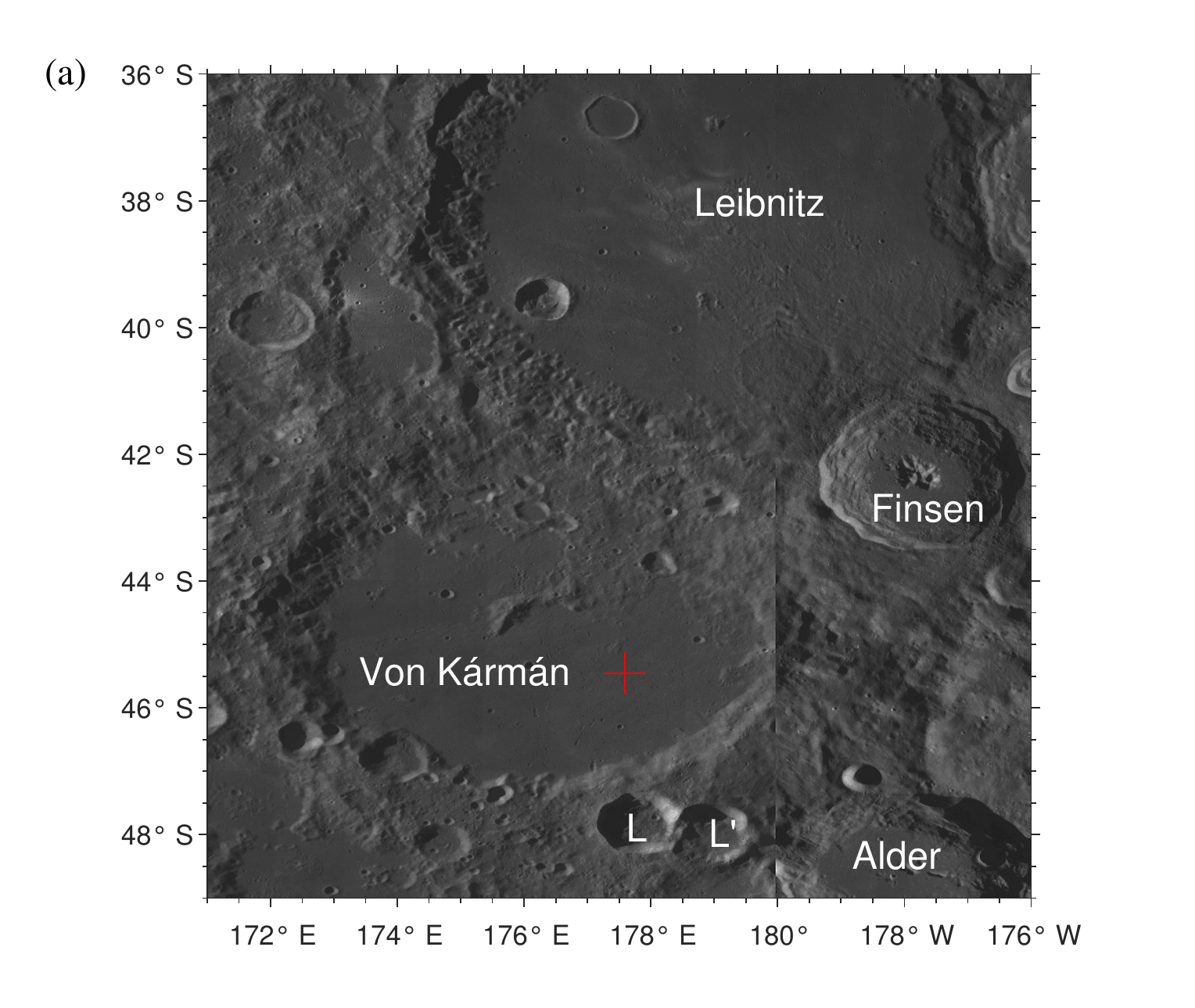}
  \includegraphics[width=8cm,angle=0]{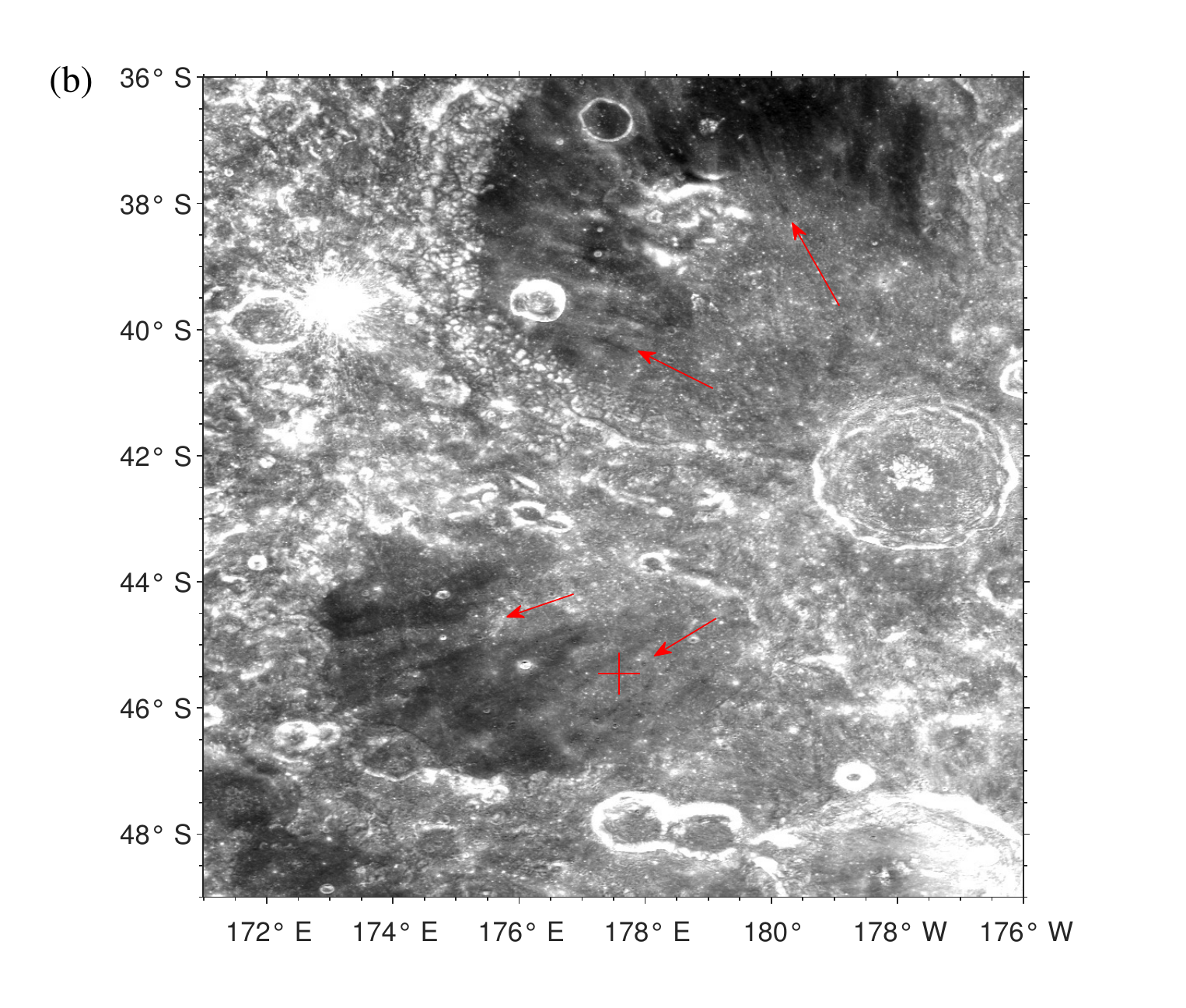}
   \centering
  \includegraphics[width=8.5cm,angle=0]{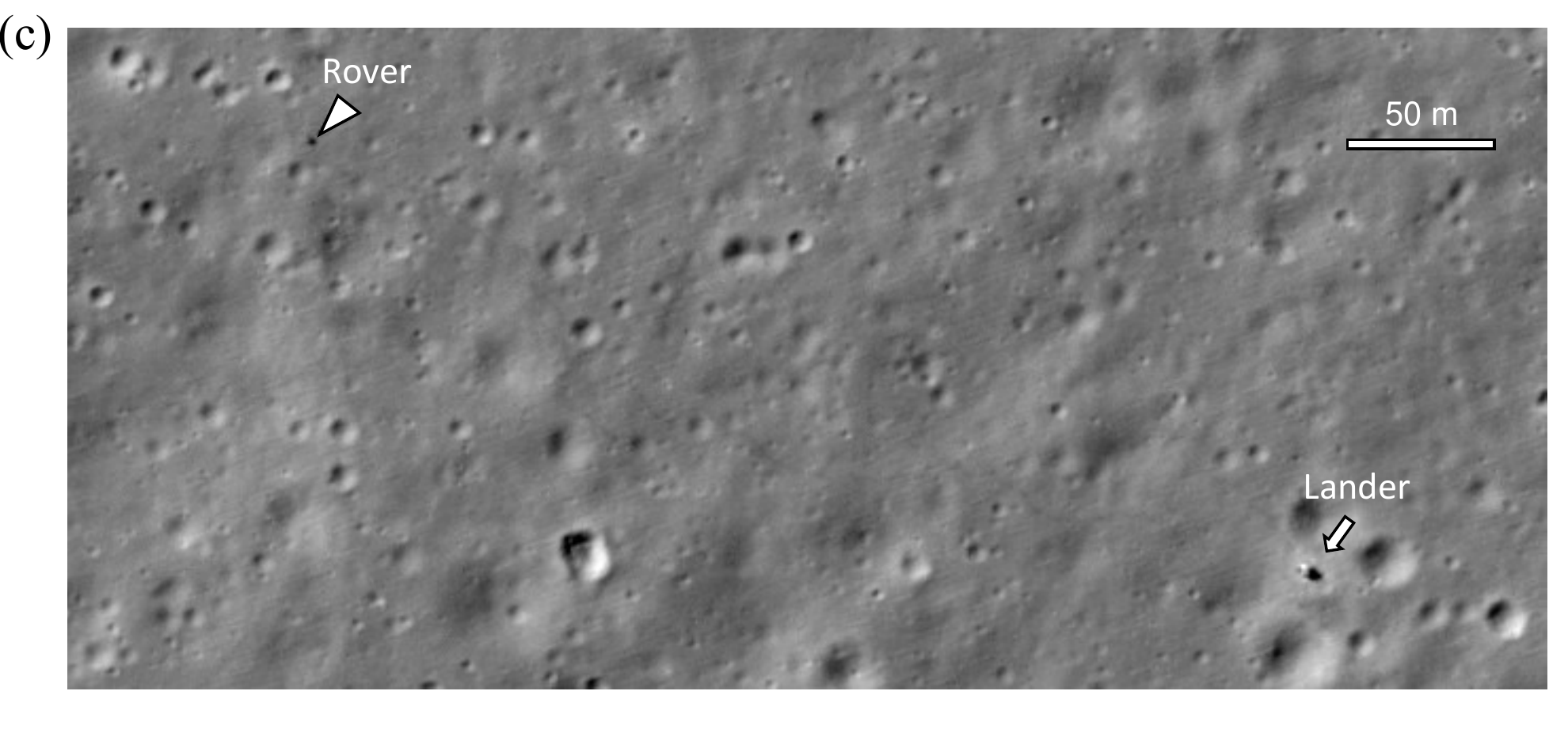}
   \caption{(a) LRO WAC image mosaic of Von Kármán crater and adjacent region. The CE-4 landing site is marked by a red cross. (b) LRO WAC mosaic at 415 nm after photometric normalization. The red arrows indicate the direction of bright ejecta from Finsen crater. (c) The CE-4 lander and Yutu-2 rover captured by LRO NAC in September, 2020. Rover tracks are faintly visible between the lander and rover.}
              \label{Fig1}%
    \end{figure}

The dielectric properties of the lunar regolith are primarily characterized by permittivity (the real component of the complex permittivity) and dielectric loss. The permittivity is used to quantify the electric polarizability of a dielectric medium. Dielectric loss (parameterized in terms of the loss tangent) quantifies a dielectric material's inherent dissipation of electromagnetic (EM) energy (e.g., heat). The permittivity determines the propagation velocity of EM waves, while the dielectric loss constrains how deeply the EM waves can penetrate into a medium. The dielectric properties of Apollo samples have been measured in a laboratory setting \citep[e.g.,][]{Carrier,Olhoeft} and these experiments showed that the relative permittivity of lunar regolith is density-dependent, while the loss tangent is a function of both density and TiO${}_{2\ }$+ FeO content. Recently, \citet{Fa} reanalyzed the Apollo samples data and suggested that the loss tangent has a significant dependence on titanium but not iron. Using LRO Diviner surface temperatures and the microwave radiometer (MRM) on board the Chang'E-2 (CE-2) orbiter, \citet{Feng2020} and the companion \citet{Siegler} produced global maps of loss tangent at 3, 7.8, 19.35, and 37 GHz. These maps not only showed a positive correlation between the loss tangent and titanium abundance, but also unveiled differences in highland and mare material as a function of frequency.

The CE-4 landing site (45.456$\mathrm{{}^\circ}$S, 177.588$\mathrm{{}^\circ}$E) is located on the eastern floor of the Von K\'{a}rm\'{a}n crater, in the South Pole-Aitken (SPA) basin --- the largest impact basin on the Moon. The Von K\'{a}rm\'{a}n crater formed in the Nectarian epoch with its floor flooded by low-albedo mare basalt in the late Imbrian age \citep{Pasckert}. The subsurface structure of the CE-4 landing site preserves a record of the sequence of major impact events of this region. Notably, there are several younger large (>25 km) impact craters around the Von K\'{a}rm\'{a}n that may contribute ejecta materials to the local regolith in the Von K\'{a}rm\'{a}n: Leibnitz, Von K\'{a}rm\'{a}n L, Alder, and Finsen craters \citep{Wilhelms1979}. The eastern part of the mare surface where CE-4 landed is covered by bright ejecta. Mineralogy and morphology analysis suggest that the Finsen crater is the primary source of this bright ejecta \citep{Hunang}. Figure \ref{Fig1}a shows the mosaic of the LRO Wide Angle Camera (WAC) for the regional vicinity of the Von K\'{a}rm\'{a}n crater. Figure \ref{Fig1}b gives the WAC mosaic at 415 nm after photometric normalization, showing that bright ejecta from Finsen crater covers a large portion of Von K\'{a}rm\'{a}n crater and Leibnitz crater. Figure \ref{Fig1}c shows an LRO Narrow Angle Cameras (NAC) image of the landing site captured in September, 2020. The CE-4 lander and Yutu-2 rover can be seen from the NAC image.

\section{Data and method}

\begin{figure*}
 \includegraphics[width=\hsize]{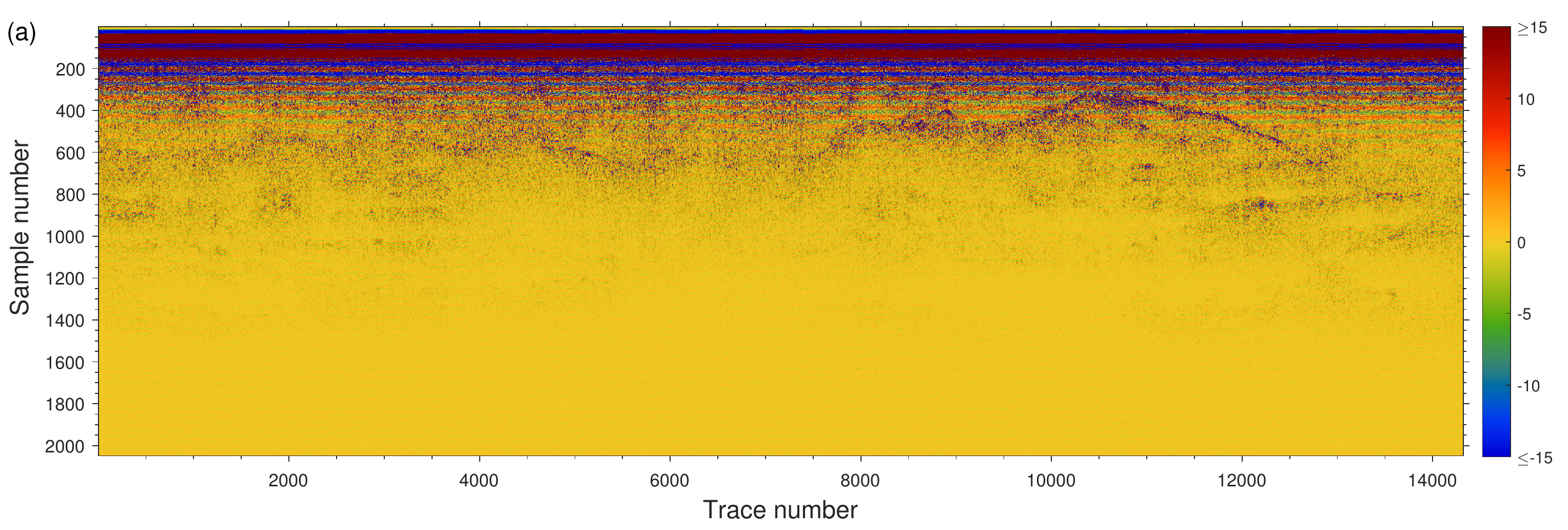}
\includegraphics[width=\hsize]{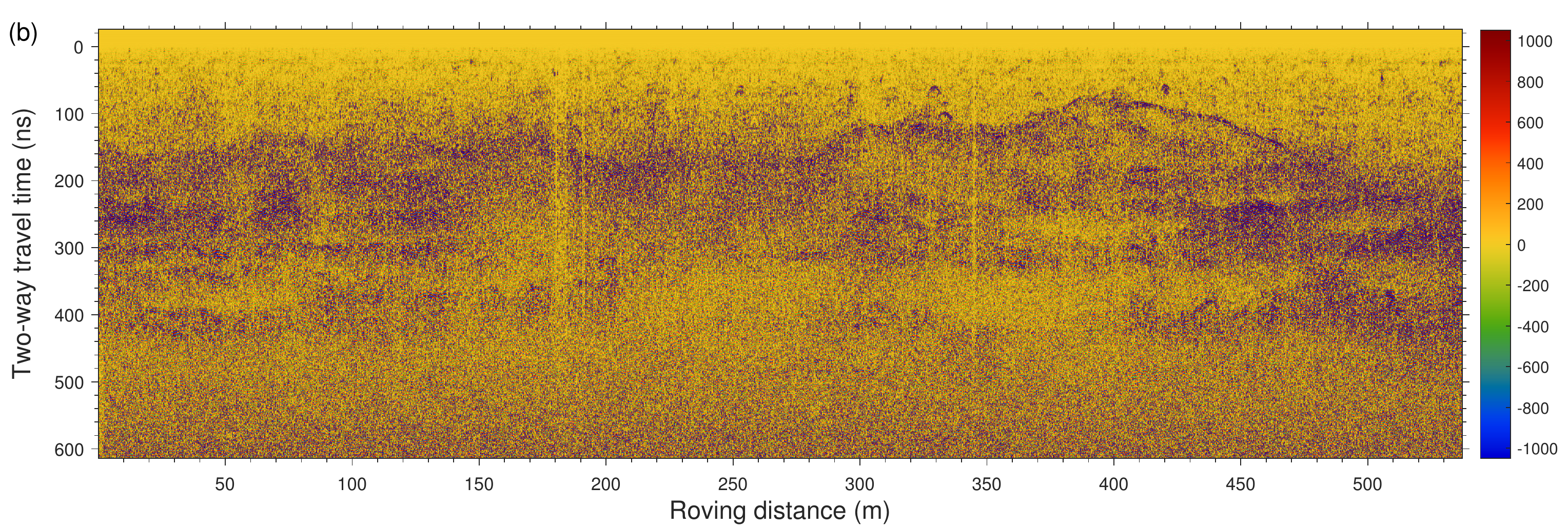}
   \caption{(a) Raw data obtained by CE-4 LPR second channel (500 MHz) from January 2019 to September 2020. The sampling interval is 0.3125 ns and the average trace interval is 3.75 cm. The colors represent relative electric field strength of echoes received by antenna. The radargram is shown in a dynamic range of $-$15 to 15 to reveal more details in deeper regolith. (b) LPR data after bandpass filtering, background removal, time-zero correction, and spherical exponential compensation.}
              \label{Fig2}%
    \end{figure*}

We processed the published LPR second channel data obtained from January 2019 to September 2020. The CE-4 LPR is a common-offset radar and its data processing includes some regular steps widely used in GPR, such as time-zero correction, bandpass filtering, and background removal \citep{jol} (similar to the data processing of CE-3 LPR). The raw data is shown in Fig. \ref{Fig2}a, with all the repeating traces deleted. In contrast to CE-3 LPR data, a continuous subsurface interface is visible in the CE-4 raw data without any processing (between 400--800 samples). A finite impulse response (FIR) filter with a pass range of 250--750 MHz was used for the bandpass filtering. To remove the background, we take the mean of all traces and subtract it from each trace in each section. Strong "ringing" and reverberation in the shallow near-surface layer are removed through the background removal. A spherical exponential compensation (SEC) gain is then implemented to correct the signal amplitude for loss of energy due to the geometrical spreading effect. After the data processing, the subsurface structure in deep regolith with more details are revealed (Fig. \ref{Fig2}b). Multiple interfaces are able to be distinguished, with the deepest one at the two-way travel time larger than 400 ns.

\begin{figure*}
    \centering
   \includegraphics[width=4cm,angle=0]{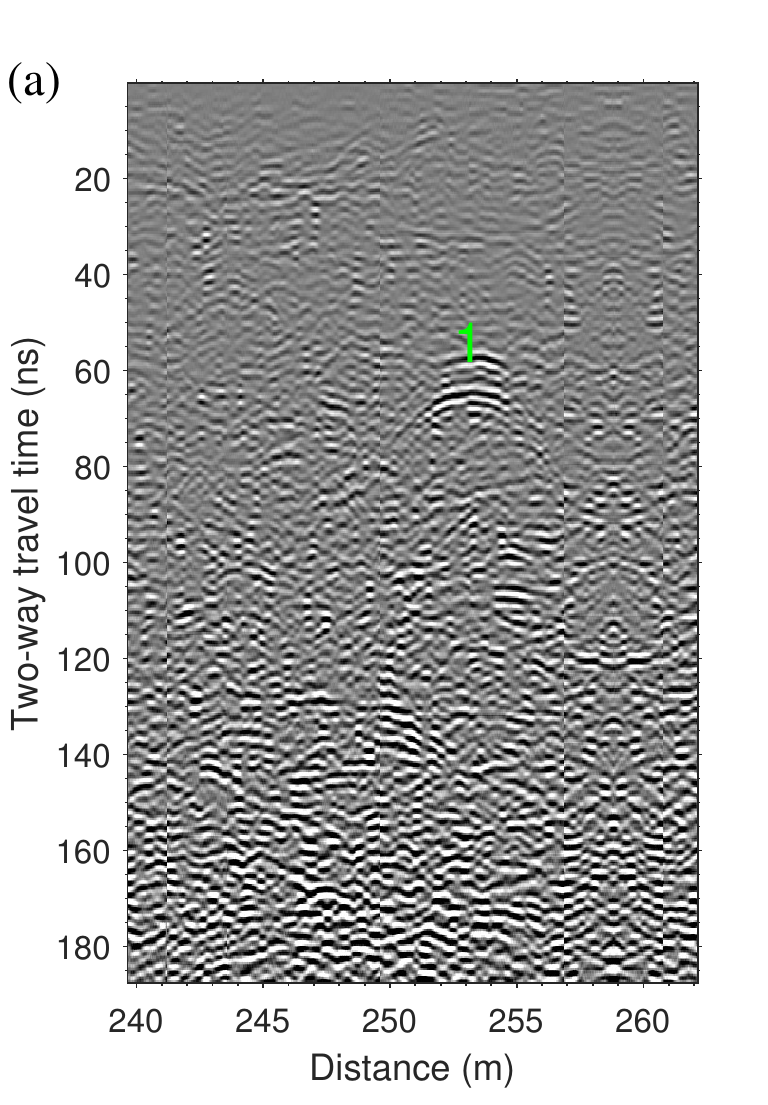}
   \includegraphics[width=4cm,angle=0]{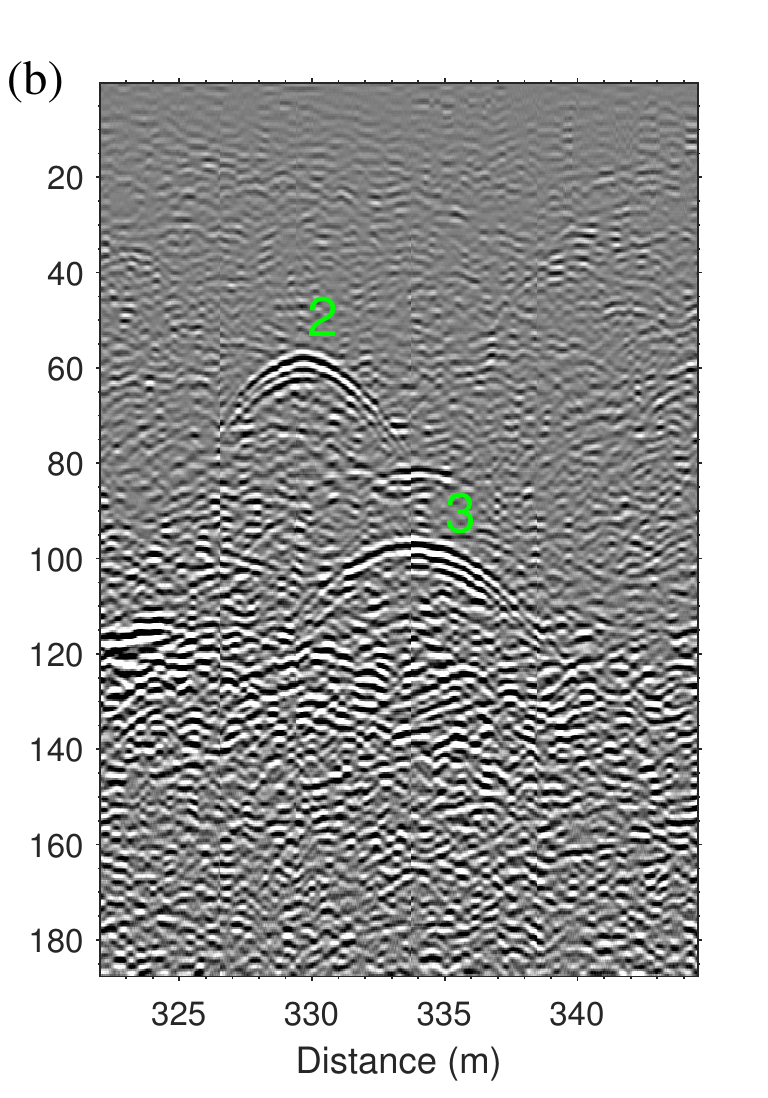}
   \includegraphics[width=4cm,angle=0]{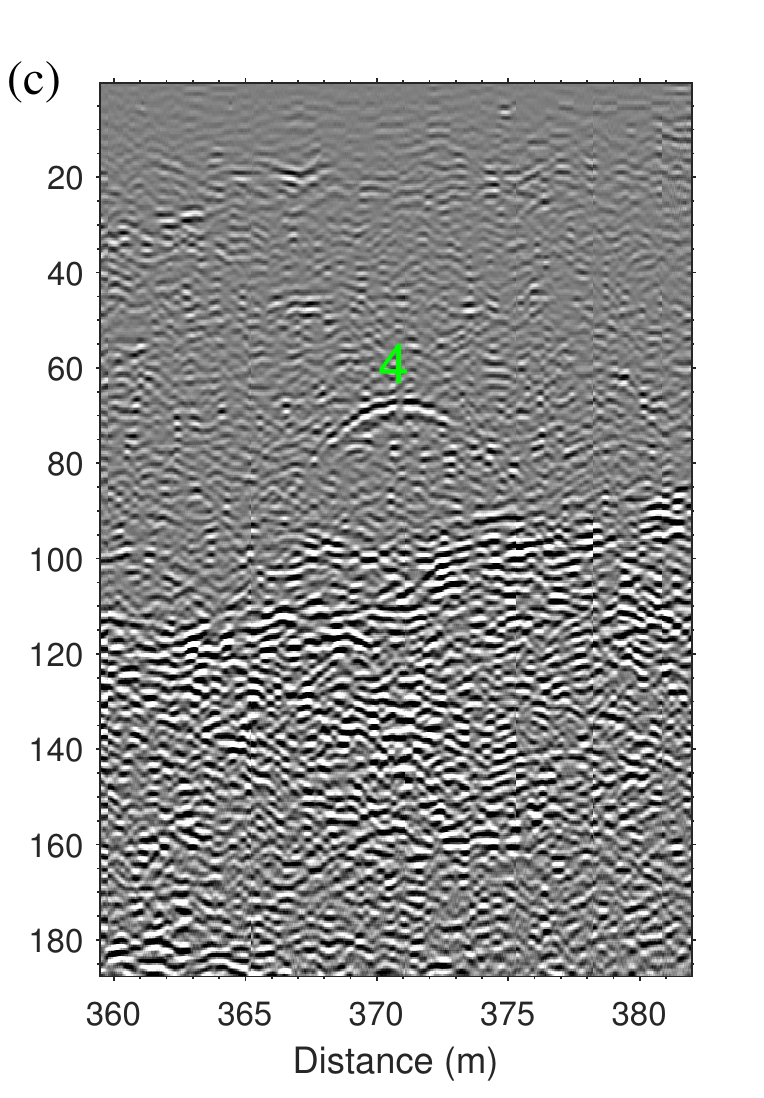}
    \includegraphics[width=4cm,angle=0]{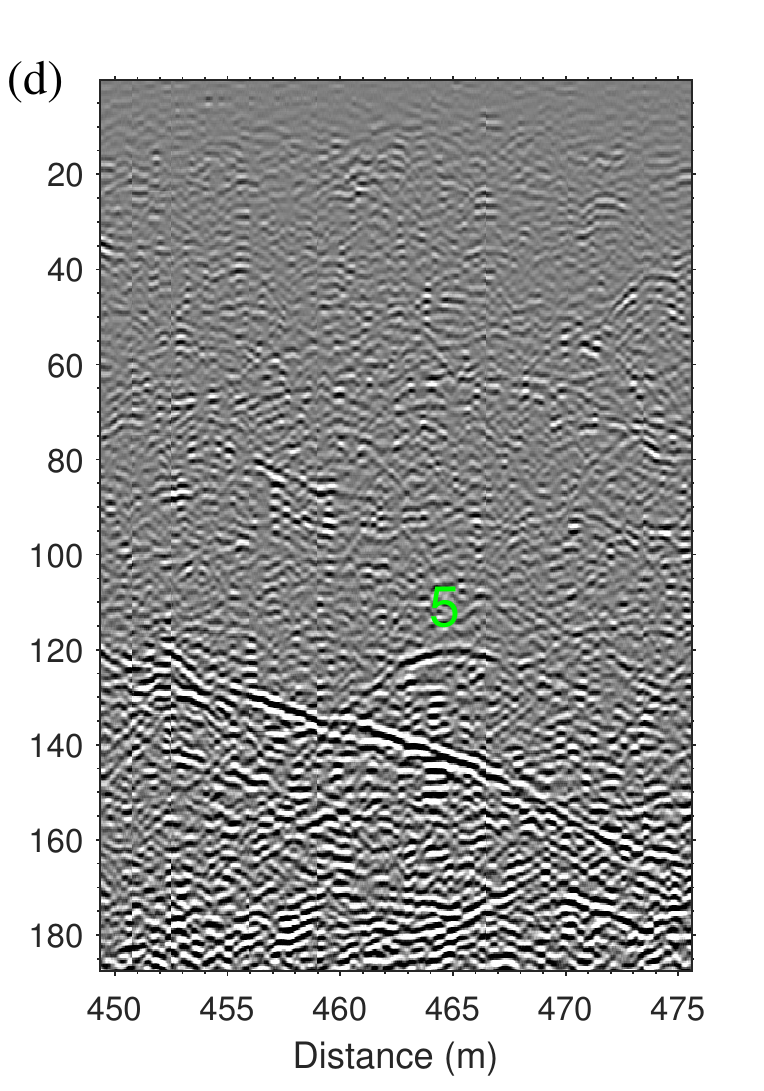}
   \caption{Noticeable hyperbolas caused by subsurface rocks. The travel velocities of radar signal estimated by curve fitting from these hyperbolas are (1) 15.97 $\pm$ 0.48, (2) 15.63 $\pm$ 0.30, (3) 16.38 $\pm$ 0.30, (4) 17.08 $\pm$ 0.36, and (5) 16.50 $\pm$ 0.31 cm/ns.}
              \label{Fig3}%
    \end{figure*}

The permittivity $\varepsilon $ of lunar regolith can be computed from velocity $v$ of EM waves traveling in the regolith as $\varepsilon = {\left(c/v\right)}^2$, where $c$ is the light speed in the vacuum. We employ hyperbolic velocity analysis \citep{jol} in our study because buried targets beneath the surface usually reflect as hyperbola patterns in GPR radargrams. The same method has been successfully used on CE-3 LPR data \citep{Feng2017,Lai2016}. The equation relating the two-way travel time, $t$, of the reflection from a target and the radar position, $x$, can be expressed in the hyperbolic form:

   \begin{equation}
      \frac{t^{2}}{t_{0}^{2}} - \frac{\left( {x - x_{0}} \right)^{2}}{\left( {vt_{0}/2} \right)^{2}} = 1,
      \label{equation1}%
   \end{equation}
   
where $x_0$ and $t_0$ are the horizontal and vertical position of the subsurface target in the radargram ($x$-$t$ plane), respectively, and $v$ and $t_0$ determine the shape of the hyperbolas (or the angles between their asymptotes). By using a least-squares method to fit the hyperbolic form of the observed data, the travel speed $v$ and the target's position ${(x}_0,\ t_0)$ can be computed. To avoid incorrect recognitions, we selected the twelve clearest hyperbolas from the LPR radargram. Figure \ref{Fig3} shows examples of these hyperbolas. These pronounced curves represent strong echoes from subsurface isolated rocks. 

The decay of a radar signal through a medium is governed by geometric spreading loss, scattering loss, and dielectric loss. At the maximum penetration depth, the radar signal decays to an undetectable level and is lost in the system noise. Here, we use the radar equation and penetration depth of LPR in the landing site to estimate the average attenuation in the regolith. The basic radar range equation for GPR \citep[e.g.,][]{Annan,Hamran,Skolnik} is: $\frac{P_{\mathrm{r}}}{P_{\mathrm{t}}}=\frac{G_{\mathrm{r}}G_{\mathrm{t}}\lambda ^2\sigma }{(4\pi )^3R^4}e^{-4\alpha R}$, where\textit{ }$P_{\mathrm{r}}$ is the received power, $P_{\mathrm{t}}$ is the transmitted power, $G_{\mathrm{r}}$ is the gain of receiving antenna, $G_{\mathrm{t}}$ is the gain of transmitting antenna, and $\lambda $ is the wavelength of EM signal in the medium. $R$ is the distance between the target and radar,\textit{ }$\alpha $ is the attenuation constant in the medium, and $\sigma $ is the backscatter radar cross-section (RCS) of the subsurface target or interface. According to \citet{Noon} we rewrite the radar equation for CE-4 LPR as:

   \begin{equation}
      G_{\mathrm{sys}} = \frac{P_{\mathrm{t}}G_{\mathrm{r}}G_{\mathrm{t}}}{P_{\mathrm{min}}} = \left\lbrack {\frac{\mathrm{T}^{2}\lambda^{2}\sigma}{\left( 4\pi)^{3}R_{\mathrm{max}}^{4} \right.}e^{- 4\alpha R_{\mathrm{max}}}} \right\rbrack^{- 1},
      \label{equation2}%
   \end{equation}
   
where $G_{\mathrm{sys}}$ is the system gain of LPR, which is 133.3 dB \citep{Zhanghongbo}; $P_{\mathrm{min}}$ is the minimum detectable signal power; $\mathrm{T}$ is the transmissivity at the vacuum-regolith interface; and $R_{\mathrm{max}}$ is the maximum penetration depth. We assume the attenuation $\alpha $ is dominated by an effective dielectric loss, which includes the scattering effect. The laboratory measurements of Apollo samples at 450 MHz indicate that lunar regolith has a loss tangent, $\mathrm{tan}\,\delta $, smaller than 0.03 \citep{Carrier}. When the $\mathrm{tan}\,\delta \ll 1$, $\alpha\approx\pi\frac{1}{\lambda}\mathrm{tan}\,\delta $. In this study, since subsurface layers at the CE-4 landing site are detected (shown in Fig. \ref{Fig2}b), we consider $\sigma $ as the RCS of a planar reflector. Given that the roughness of subsurface interfaces relative to the LPR center wavelength is greater than 0.5/$\pi$, these interfaces are considered as rough planar reflectors, and their cross-section can be taken as the area of the "first Fresnel zone," with $\sigma =\mathrm{\Gamma }\pi \lambda R/2$ \citep[e.g.,][]{Annan,Cook,Grimm,Noon}. Therefore, the average loss tangent of regolith could be calculated from the maximum signal penetration distance by:

   \begin{equation}
    \mathrm{tan}\,\delta = \frac{\lambda}{4\pi R_{\mathrm{max}}}{\ln\left\lbrack \frac{\mathrm{T}^{2}\lambda^{3}\Gamma G_{\mathrm{sys}}}{128\pi^{2}R_{\mathrm{max}}^{3}} \right\rbrack}.
    \label{equation3}%
   \end{equation}
   
Here, $R_{\mathrm{max}}$ can be replaced by maximum travel time $t_{\mathrm{max}}$ in the form of $R_{\mathrm{max}}={f\lambda t_{\mathrm{max}}}/{2}$, where $f$ is the signal frequency, $\lambda$ is the signal wavelength in regolith ($\lambda = {\lambda_{0}/\sqrt{\varepsilon}}$, where $\lambda_{0}$ is the signal wavelength in vacuum). Equation \eqref{equation3} then becomes:

   \begin{equation}
    \mathrm{tan}\,\delta = \frac{1}{2\pi{ft}_{\mathrm{max}}}{\ln\left\lbrack \frac{\mathrm{T}^{2}\Gamma G_{\mathrm{sys}}}{16\pi^{2}f^{3}t_{\mathrm{max}}^{3}} \right\rbrack}.
    \label{equation4}%
   \end{equation}
   
\begin{figure}
    \centering
   \includegraphics[width=8.5cm,angle=0]{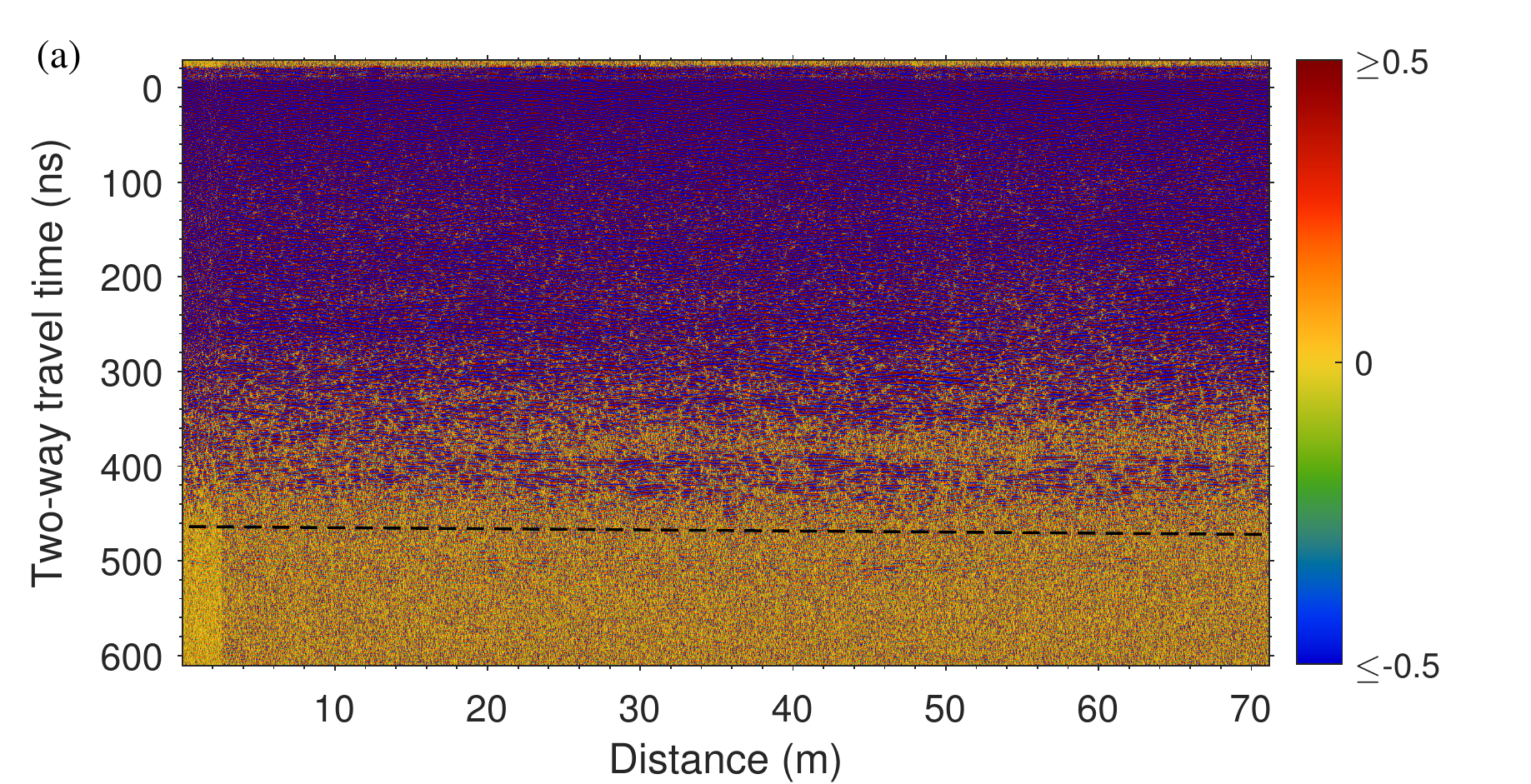}
   \includegraphics[width=8.5cm,angle=0]{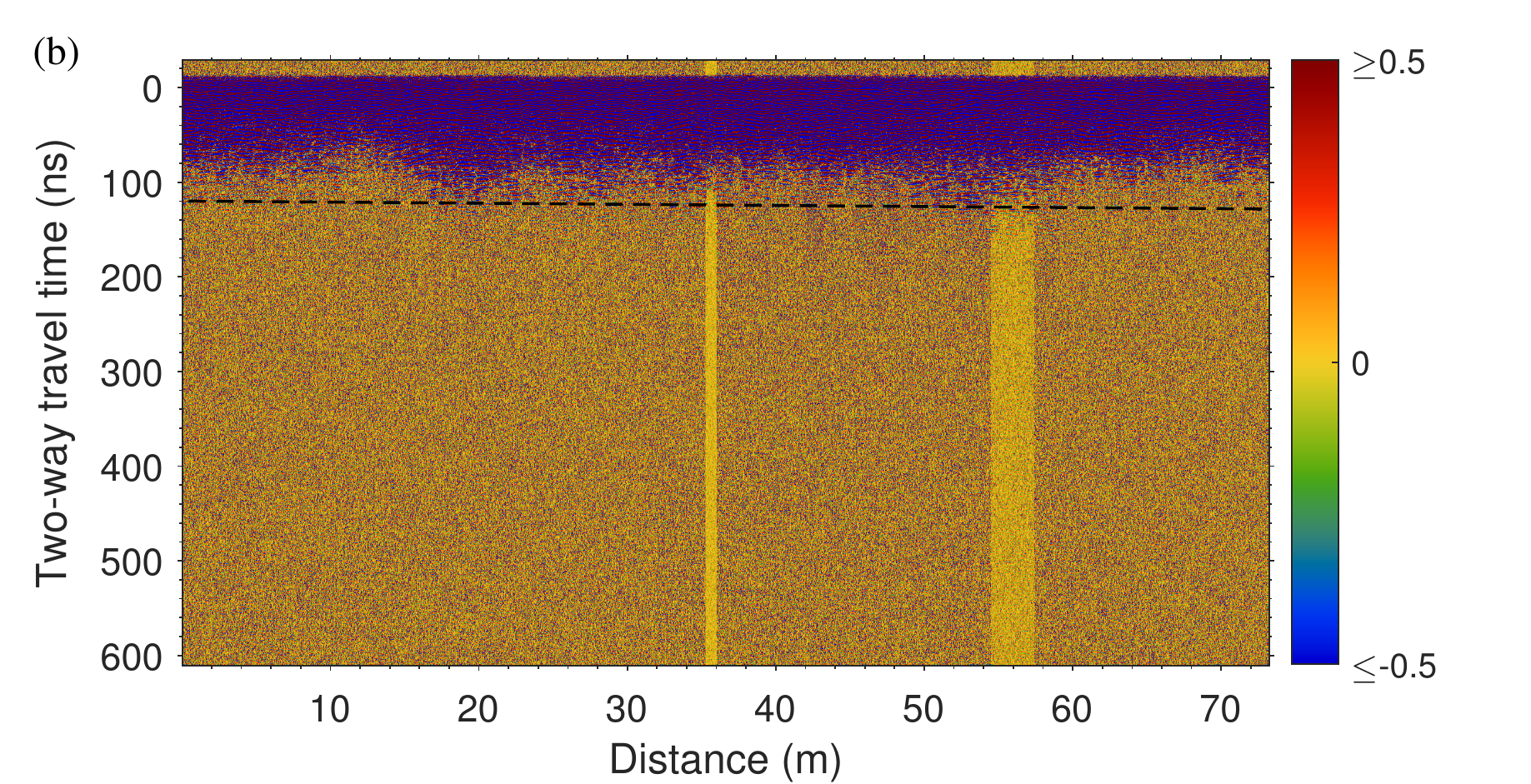}
   \caption{Boundary between signals and noise in the LPR data. (a) A subset of CE-4 LPR radargram shown in a small dynamic range (b) compared with that of CE-3 LPR. The dashed lines indicate the maximum penetration time of the radar signals.}
              \label{Fig4}%
    \end{figure}

For low-loss mediums, the reflectivity $\mathrm{\Gamma }$ of the substrate can be calculated from $\mathrm{\Gamma }={\left|\frac{\sqrt{\varepsilon }-\sqrt{{\varepsilon }_1}}{\sqrt{\varepsilon }+\sqrt{{\varepsilon }_1}}\right|}^2$, where\textit{ }$\varepsilon $ and ${\varepsilon }_1$ are the permittivity of regolith and rocks (we assume ${\varepsilon }_1$= 7 in this study \citep{Carrier}). The transmissivity of the regolith surface is determined by $\mathrm{T}=\frac{4\sqrt{\varepsilon }\sqrt{{\varepsilon }_0}}{{\left|\sqrt{\varepsilon }+\sqrt{{\varepsilon }_0}\right|}^2}$, where ${\varepsilon }_0$ is the permittivity of the vacuum. In Figure \ref{Fig4}, we compare the radargram of CE-4 LPR with that of CE-3 LPR, both of which are shown in a very small dynamic range ($-$0.5 to 0.5) to display the boundary between signals and noise. It can be seen that the maximum penetration time at the CE-4 landing site is roughly four times longer than that at the CE-3 landing site.

To better interpret the subsurface features in the deposits and convert the travel time to penetration depth, an imaging method was applied to the radargram. The algorithm we used is the Kirchhoff migration. In migration, observed reflections are moved from the apparent locations back to their real spatial locations. A hyperbola in the radargram will be ``refocused'' to a target's position. Kirchhoff migration is based on an integral solution of the scalar wave equation. According to \citet{Schneider} and \citet{Ozdemir}, in a 2-D situation the far-field approximation of Kirchhoff integral for LPR can be expressed as:

   \begin{equation}
U\left( x_0,z_0 \right) = \frac{1}{2\pi}{\int{\left\lbrack \frac{\mathrm{cos}\,\theta}{v_{m}R}\frac{\partial}{\partial t}U\left( x,z,t \right) \right\rbrack dx}}
   \label{equation5}%
   ,\end{equation}

where $U(x_0,z_0)$ is the wave field at a subsurface scatter point $(x_0,z_0)$, $U(x,z,t)$ is the wave field observed by LPR at surface position ($x,z$) at time two-way travel time $t$, $\theta $ is the refraction angle of wave field in the medium, $R$ is the propagation distance of radar signal from the point to the observation position, and $v_m$ is half of the travel speed in the regolith.

\section{Result}
\subsection{Permittivity and dielectric loss}

   \begin{figure*}
    \centering
   \includegraphics[width=6.5cm,angle=0]{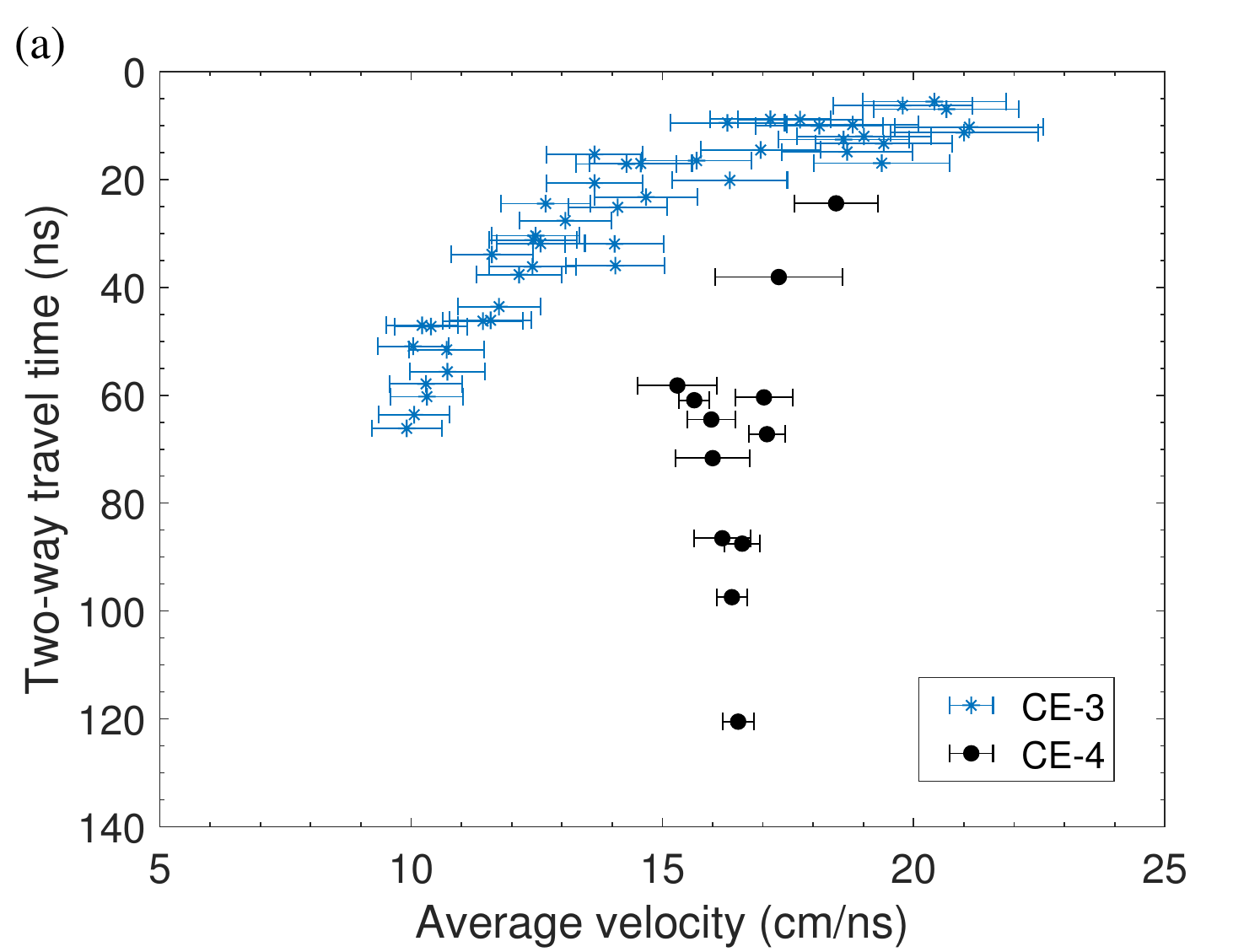}
  \includegraphics[width=6.5cm,angle=0]{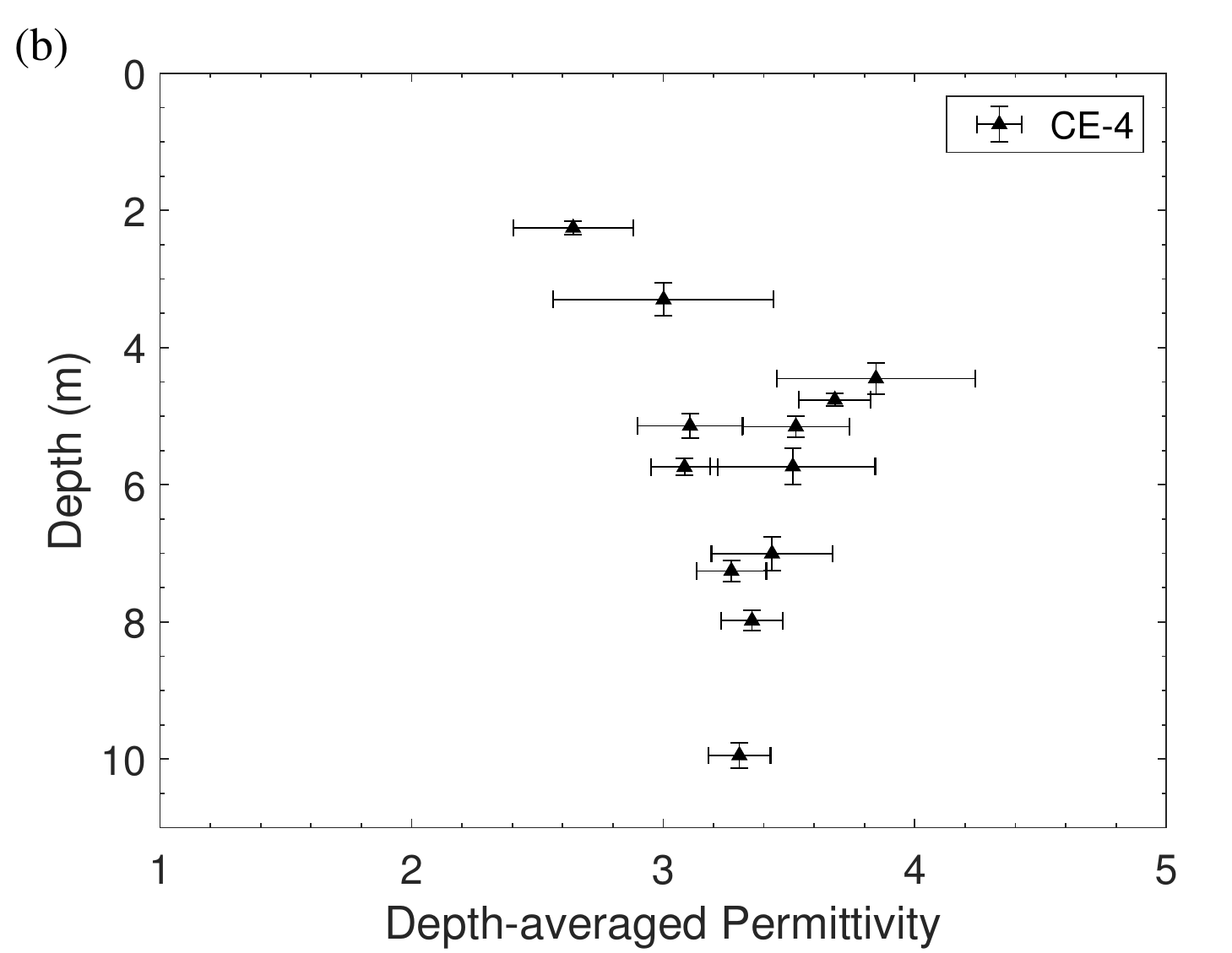}
   \includegraphics[width=6.5cm,angle=0]{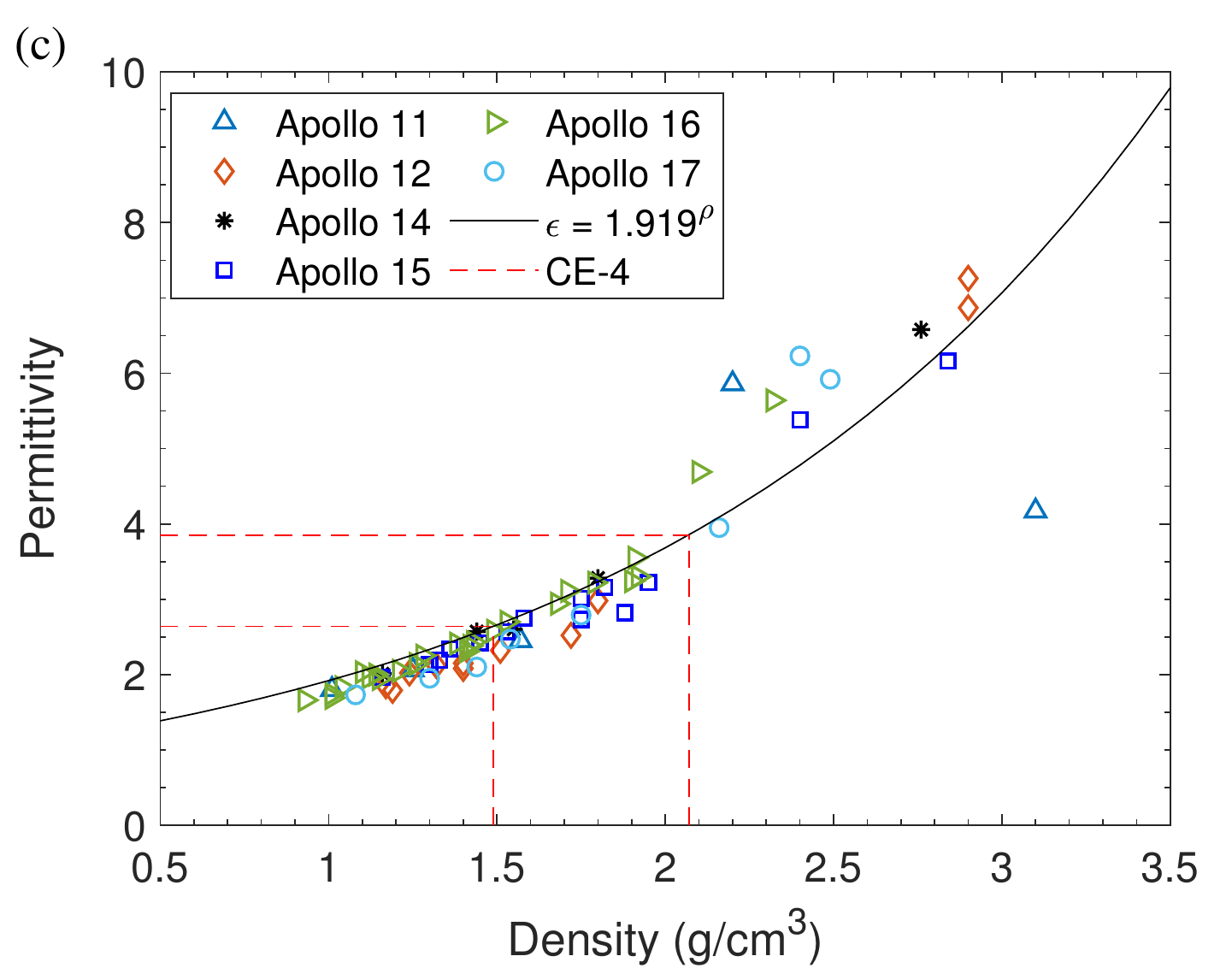}
    \includegraphics[width=6.5cm,angle=0]{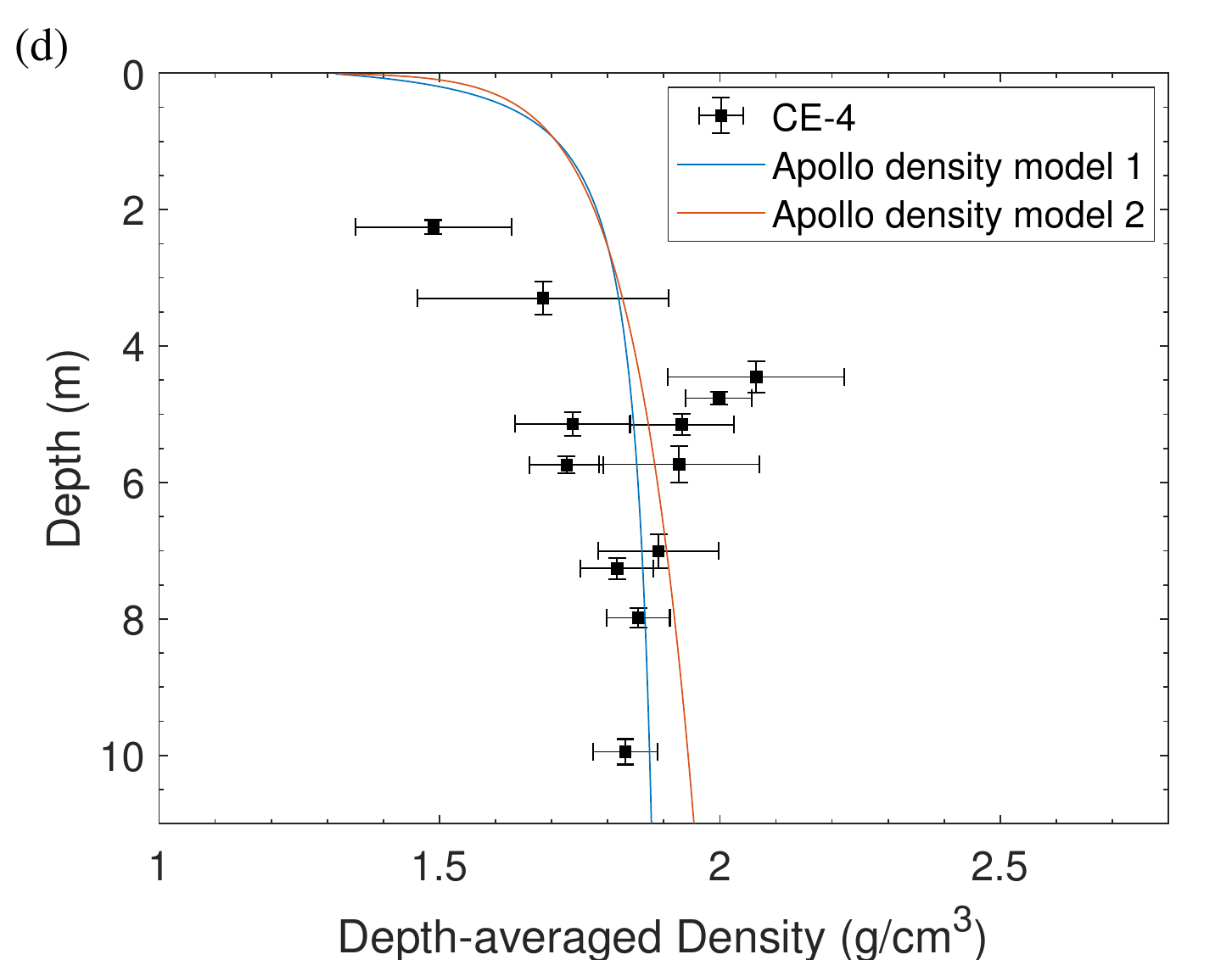}
   \caption{Permittivity and density of lunar regolith estimated from LPR data. (a) Average propagation velocity of radar signal estimated from hyperbolas as a function of two-way travel time. (b) Depth-averaged permittivity of regolith at CE-4 landing site versus depth. (c) Permittivity of Apollo soil samples at 450 MHz as a function of density \citep{Olhoeft}. The dashed lines constrain the CE-4 permittivity range and the corresponding density range. (d) Inverted depth-averaged bulk density of regolith as a function of depth. The Apollo density models come from \citet{Carrier}}
              \label{Fig5}%
    \end{figure*}
 
As shown in Fig. \ref{Fig5}a, the average EM velocity in the regolith at the landing site ranges from 15.30 to 18.46 cm/ns. The EM velocity at the CE-3 landing site \citep{Feng2017} is also plotted in the figure for comparison.  Accordingly, the depth-averaged permittivity of CE-4 regolith varies from 2.64 to 3.85 (Fig. \ref{Fig5}b), consequently falling within the range constrained by Apollo soil samples with a density lower than 2.1 g/cm${}^{3}$ \citep{Olhoeft} (Fig. \ref{Fig5}c). Because we did not find any clear hyperbolas before 20 ns in the radargram, the shallowest point in Fig. \ref{Fig5}b for which we can calculate the dielectric constant is at a depth of 2 m. Generally, the permittivity increases with depth in the upper $\mathrm{\sim}$ 4 m but remains within the range between three and four for 4 to 10 m. This implies that the local regolith (at least the first layer) consists of fine-grained soil and is much more homogeneous than the regolith at the CE-3 landing site, where the weathered material contains more rocks and the permittivity reaches six at $\mathrm{\sim}$3 m depth \citep{Feng2017,Lai2016}. \citet{Carrier} proposed that the permittivity of lunar regolith is density-dependent in the form of $\varepsilon ={1.919}^{\,\rho }$, where $\rho $ is bulk density. The depth-averaged bulk density at the CE-4 landing site derived from this equation ranges from 1.49 to 2.07 g/cm${}^{3}$ (Fig. \ref{Fig5}c) and is always greater than 1.68 g/cm${}^{3}$ below depths of 3 m. We compare the CE-4 density with the models constrained by the measurements of Apollo core samples \citep{Carrier} in Fig. \ref{Fig5}d. The blue and red lines are deduced from the Apollo density profile model, $\rho(z) =1.92\frac{z+12.2}{z+18}$ and $\rho(z)=1.39z^{\,0.056}$ \citep{Carrier}, where $z$ is the depth in cm. In the deeper regolith, the result is in good agreement with the Apollo curves. At the shallower depths, the CE-4 density is smaller than the Apollo model prediction. This may be due to the lateral inhomogeneity (e.g., variations in mineralogy or density) or because the regolith at the CE-4 site is more porous than that of the Apollo sites.

   \begin{figure*}
        \centering
        \includegraphics[width=6.5cm,angle=0]{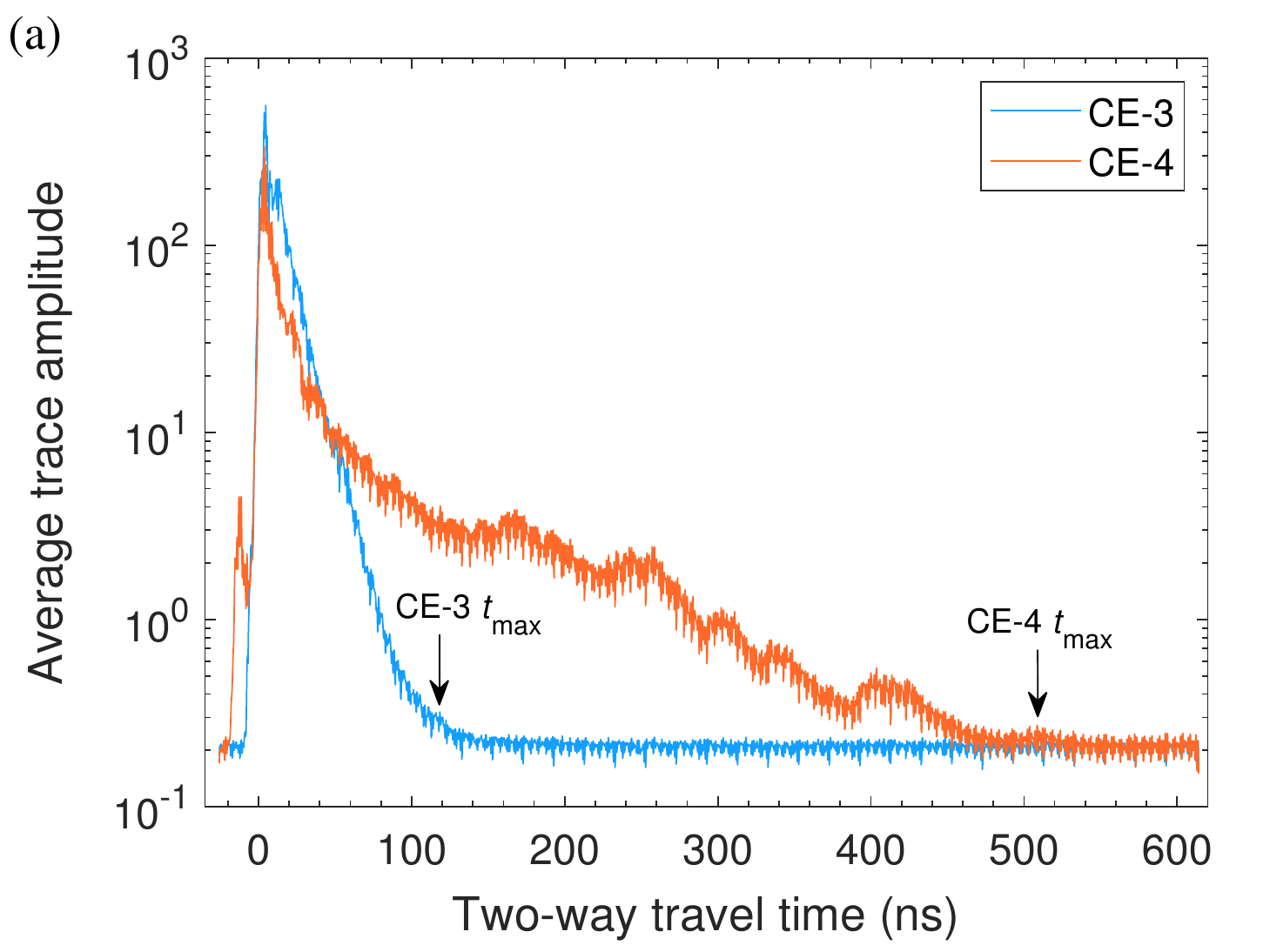}
        \includegraphics[width=6.5cm,angle=0]{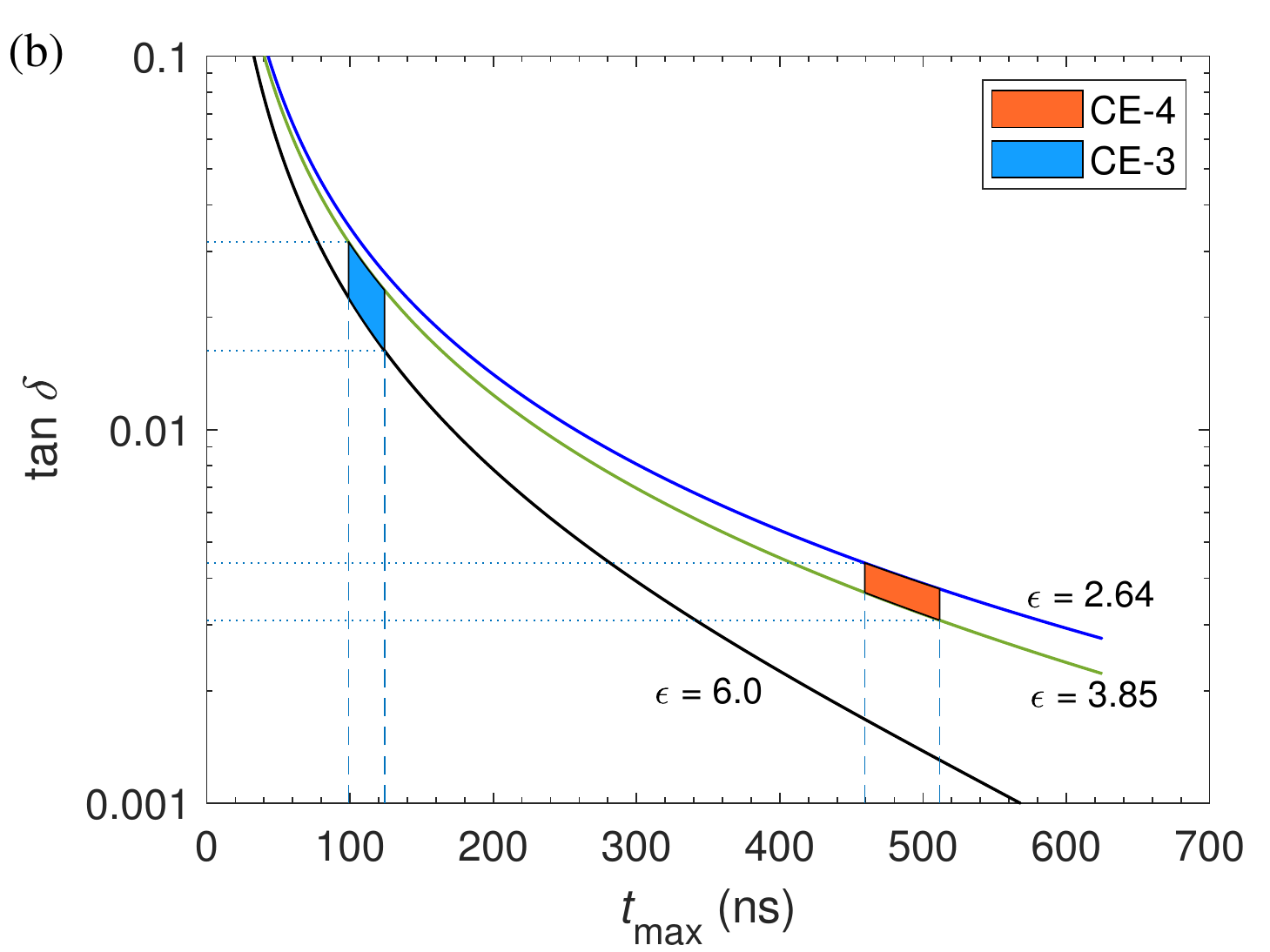}
        \includegraphics[width=6.5cm,angle=0]{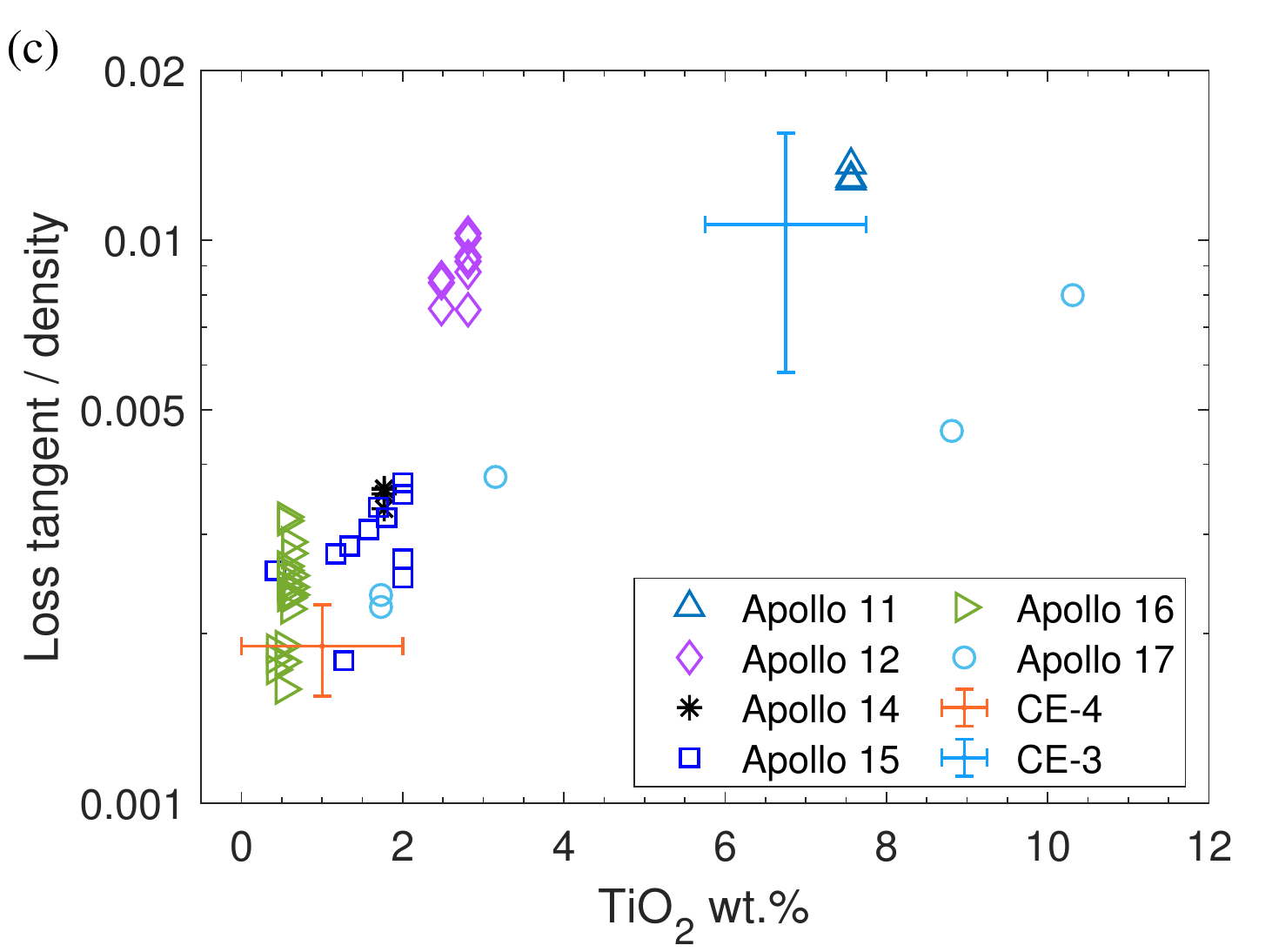}
        \includegraphics[width=6.5cm,angle=0]{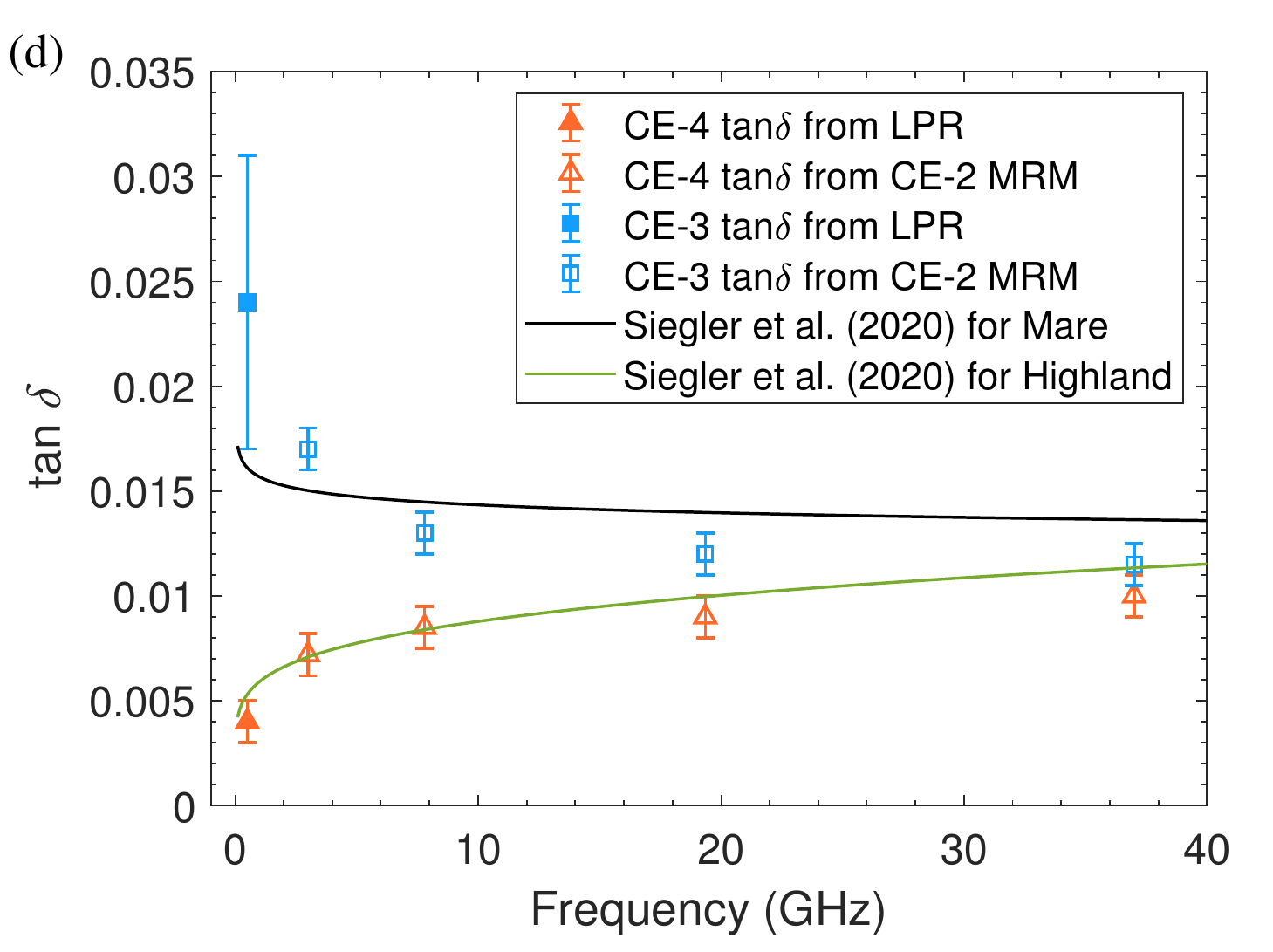}
        \caption{Dielectric loss of lunar regolith estimated from LRP data. (a) Average LPR trace amplitudes from 100 m of CE-3 data and 100 m of CE-4 data, used to assess signal decay in the lunar subsurface. The position of time-zero is corrected according to the largest amplitude. (b) The loss tangent of lunar regolith as a function of maximum penetration time. (c) The density normalized loss tangent as a function of TiO${}_{2}$ weight percentage. The Apollo data points only represent measurements at 450 MHz. (d) The loss tangent at CE-3 and CE-4 landing sites as a function of frequency. The data at 3, 7.8, 19.35, and 37 GHz are taken from \citet{Siegler}.}
        \label{Fig6}%
\end{figure*}
   
To examine the maximum penetration depth of the LPR along the Yutu-2 rover's traverse, this study uses the average trace amplitude (ATA) plot of the radar echoes. Typically, ATA plots are  used in GPR data analysis to determine the signal penetration (attenuation) from the overall decay of the ATA curve \citep[e.g.,][]{Butler,Pettinelli}. Here, we compressed the radargram after bandpass filtering and background removal into a single trace by taking the mean absolute value over all traces at each time to produce the ATA plot. Figure \ref{Fig6}a gives the ATA curve of a 100-m subsets of CE-4 radar cross-section. Given that the radar systems and rover configurations of CE-4 LPR are identical to CE-3 LPR, we compare the ATA of data from both instruments. The CE-4 LPR signal reaches the noise floor after 460 ns. The small peaks on the ATA curve represent the reflections from subsurface interfaces, agreeing with the features shown in Fig. \ref{Fig2}b. The last peak at $\mathrm{\sim}$ 510 ns refers to the echoes from the deepest planar reflector LPR could detect, so 460--510 ns is considered as the range of maximum travel time $t_{\mathrm{max}}$ of the LPR signal. The loss tangent of local regolith as a function of $t_{\mathrm{max}}$ is shown in Fig. \ref{Fig6}b, wherein a penetration curve corresponds to a specific permittivity of local regolith. Penetration curves with permittivities of 2.64 and 3.85 are used as the upper and lower boundaries at the CE-4 landing site, resulting in dielectric loss tangents ranging from 0.0031--0.0044, according to the $t_{\mathrm{max}}$ range.

In contrast to the CE-4 landing site, the attenuation at the CE-3 landing site is much greater, with echoes reaching noise floor at 100--120 ns. According to the calculation from \citet{Lai2016} and \citet{Feng2017}, we used 3.85 and 6 as the upper and lower boundaries of CE-3 regolith permittivity in Fig. \ref{Fig6}b. This is meant to lead to a loss tangent falling in the range of 0.016--0.032. However, because there is no apparent planar reflector or interface in the regolith at the CE-3 landing site \citep{Feng2017,Lai2016}, Eqs. \eqref{equation3} and \eqref{equation4} are not adequately applicable and the loss tangent may be inaccurate. The calculated value may just provide an upper boundary of CE-3 $\mathrm{tan}\,\delta $, since a regolith or bedrock interface is assumed to be at the maximum penetration depth. Even so, from these estimates we can infer that the lunar regolith is more highly attenuating beneath CE-3 than under CE-4.

Figure \ref{Fig6}c compares $\mathrm{tan}\,\delta$ of CE-4 regolith with that of soil samples ($\rho $ $\mathrm{<}$ 2.1 g/cm${}^{3}$) from Apollo 11, 12, 14, 15, 16, and 17 \citep{Carrier}. The Apollo data clearly show the density normalized $\mathrm{tan}\,\delta $ is positively correlated with the TiO${}_{2}$ abundance. Due to a low TiO${}_{2}$ content (less than 2\%) \citep{Sato}, the density normalized $\mathrm{tan}\,\delta $ of CE-4 regolith is very close to the low $\mathrm{tan}\,\delta $ of highland samples brought back by Apollo 14, 15 16, and 17 missions. CE-3 site has a higher TiO${}_{2}$ content of $\mathrm{\sim}$6.7\% \citep{Sato}, and its $\mathrm{tan}\,\delta $ is close to that of mare samples from Apollo 11. Figure \ref{Fig6}d combines the results from LPRs and CE-2 microwave radiometer, showing that the $\mathrm{tan}\,\delta $ at CE-4 site increases with frequency, while at CE-3 site the $\mathrm{tan}\,\delta $ decreases with frequency. These trends are consistent with the $\mathrm{tan}\,\delta $ models proposed by \citet{Siegler} as $\mathrm{tan}\,\delta ={10}^{\,0.312\rho +f^{\,0.069}-3.79}$ for highland and $\mathrm{tan}\,\delta ={10}^{\,0.312\rho +(f^{\,-0.0025}-0.958)\times \%\mathrm{TiO_2}-2.65}$ for mare, where $f$ is frequency in GHz. Both Figs. \ref{Fig6}c and \ref{Fig6}d indicate that the dielectric behavior of CE-4 regolith is similar to that of highland materials.

\subsection{Stratigraphy}
\begin{figure*}
\centering
\includegraphics[width=\hsize]{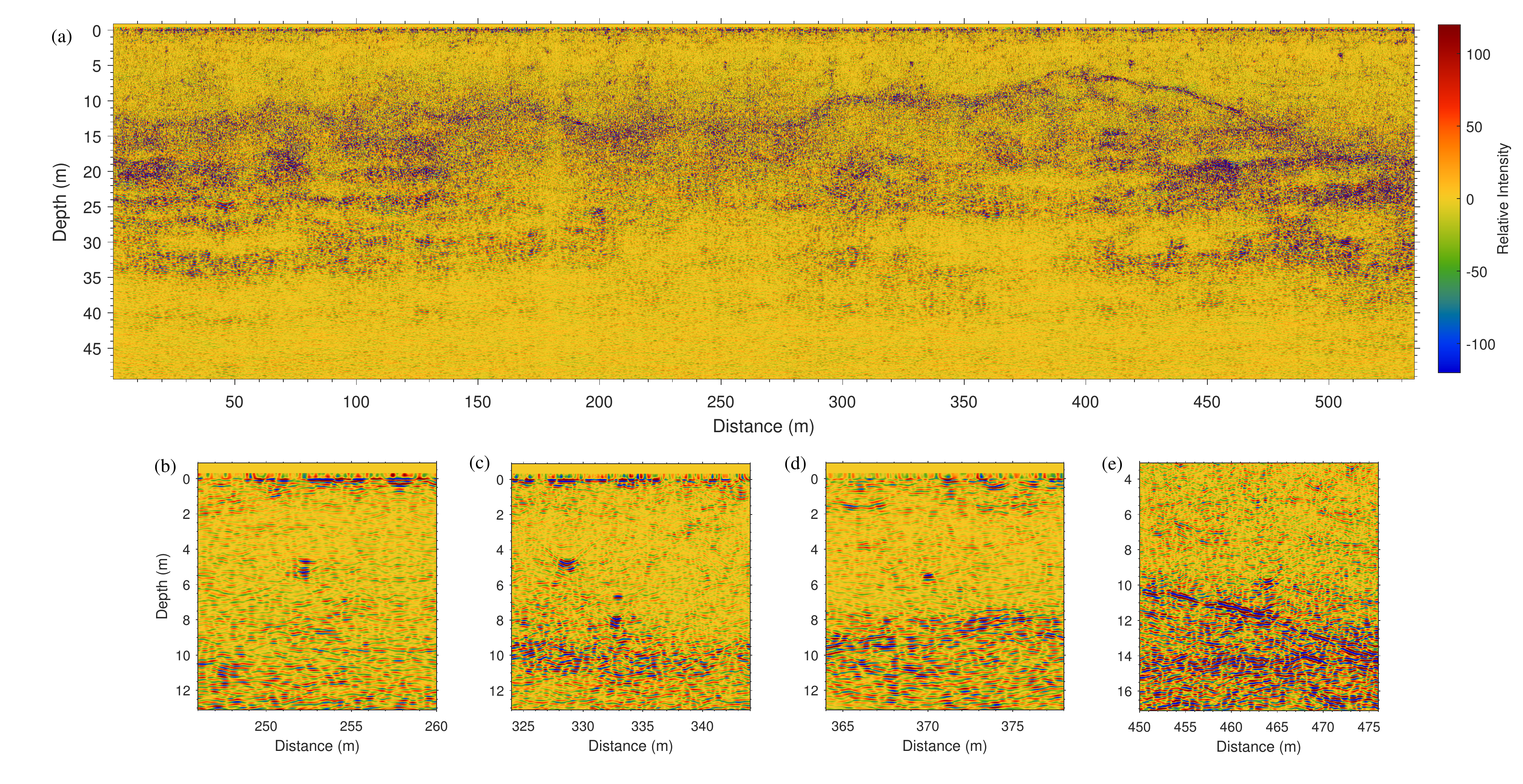}

\caption{Subsurface profile of lunar regolith at the CE-4 landing site. (a) Reconstruction of lunar regolith subsurface structure using migration. The colors represent relative intensity of reflections from subsurface objects. (b)-(e) display the migration results of the hyperbolas given in Fig. \ref{Fig3}.}
\label{Fig7}
\end{figure*}

The regolith structure of the CE-4 landing site is unveiled in the radargram after migration demonstrated in Fig. \ref{Fig7}a, with Figs. \ref{Fig7}b--\ref{Fig7}e showing that each hyperbola in Fig. \ref{Fig3} is refocused to a rock's position. The mean travel velocity below 3 m depth (16.3 cm/ns) was applied in the migration. Several planar reflectors appear and split the cross-section into multiple subsurface layers. These reflectors, at the bottom of each layer, are composed of rock fragments and fresh impact ejecta from nearby craters. Due to the continuous comminution and reworking by micrometeorite impact, the upper material in each layer is overturned and pulverized much more frequently, thus taking on a smaller and more uniform grain size. The top layer thickness is $\mathrm{\sim}$ 7--16 m with a relatively "transparent" background, indicating a mature fine grain material with low rock abundance. This is consistent with the smooth surface seen in the picture of the landing site captured by the panoramic camera on board the rover \citep{Li}. The permittivity result also shows that this fine-grained layer that lacks scatterers is relatively homogeneous along the rover's traverse. The other layers in the cross-section are thinner and have higher rock abundances, indicating a shorter period of gardening and evolution. These may be the ejecta delivered by craters smaller than the source of the first layer. The deepest reflector distinguishable is at a depth of $\mathrm{\sim}$ 40 m. Its echo is weak, continuous, and flat. This may reflect the boundary between paleoregolith and basaltic base rock.

\begin{figure*}
    \centering
   \includegraphics[width=16cm,angle=0]{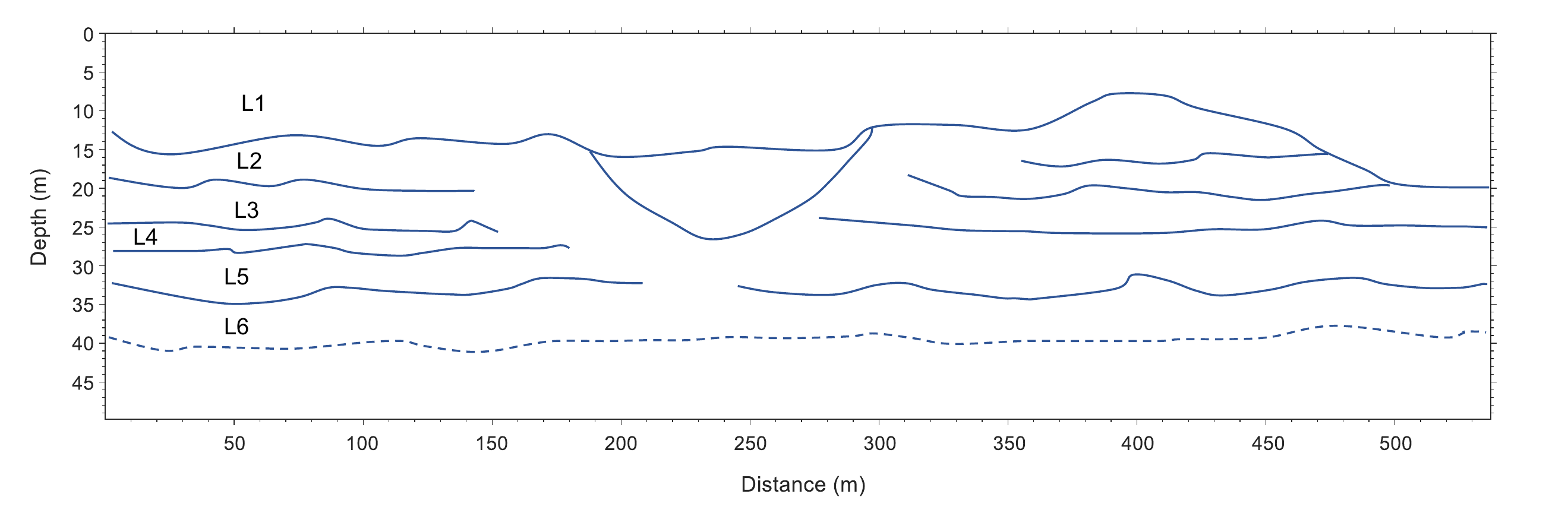}
   \caption{Stratigraphic diagram of regolith at the CE-4 landing site.}
              \label{Fig8}%
    \end{figure*}

In the middle of the radargram (between roughly 200--300 m along the traverse), subsurface reflectors or layers are disrupted by a symmetric bowl-shaped structure filled with point reflectors, as illustrated in Fig. \ref{Fig8}. It penetrates three subsurface layers with a height of $\mathrm{\sim}$ 13--15 m and a horizontal length of $\mathrm{\sim}$ 128 m on its top. The echoes beneath this structure become weaker than those on other parts of the horizontal interfaces. Using LRO NAC Digital Terrain Models, \citet{Stopar} recently found that craters with diameters less than 400 m have a depth-to-diameter (d/D) ratio ranging from 0.11 to 0.17. As the d/D ratio of this structure is $\mathrm{\sim}$ 0.10--0.12, we speculate it is the profile of an impact crater. This finding is also supported by the recent work from \citet{Zhangling}. This crater might form before the new material covered the paleoregolith surface. It is unlikely to be a secondary crater caused by the ejected debris because there is no topographic evidence on the surface of top layer (Fig. \ref{Fig1}c) and the crater was filled with fragments before it was buried. As the rock fragments filled in the crater scatter the EM signal in all directions, the echoes from deeper layers attenuate faster and the penetration depth of LPR is shallower under this carter.

The stratigraphic diagram illustrated in Fig. \ref{Fig8} displays six layers from L1 at the top to L6 at the bottom. It is inferred from the local geological context that the soil at the landing site may consist of ejecta coming from the Finsen, Von Kármán L, Von Kármán L', Alder, and Leibnitz craters. To analyze the contributions of nearby impact events to the regolith at the landing site, we estimated the thickness of distal ejecta from each crater. As reported by \citet{Mcgetchin}, the ejecta thickness model is described by $t={T(r/R)}^b$, where \textit{t} is the thickness at radial distance \textit{r} from the crater center, \textit{T} is the ejecta thickness at the rim of the crater, \textit{R} is the radius of the crater, and \textit{b} is the exponent of the power-law. \citet{Pike} pointed out \textit{R} should be the transient radius and defined \textit{T} and \textit{b} as:\textit{ }$T=0.033R$, $b=-3$. 

\begin{figure}
    \centering
   \includegraphics[width=8cm,angle=0]{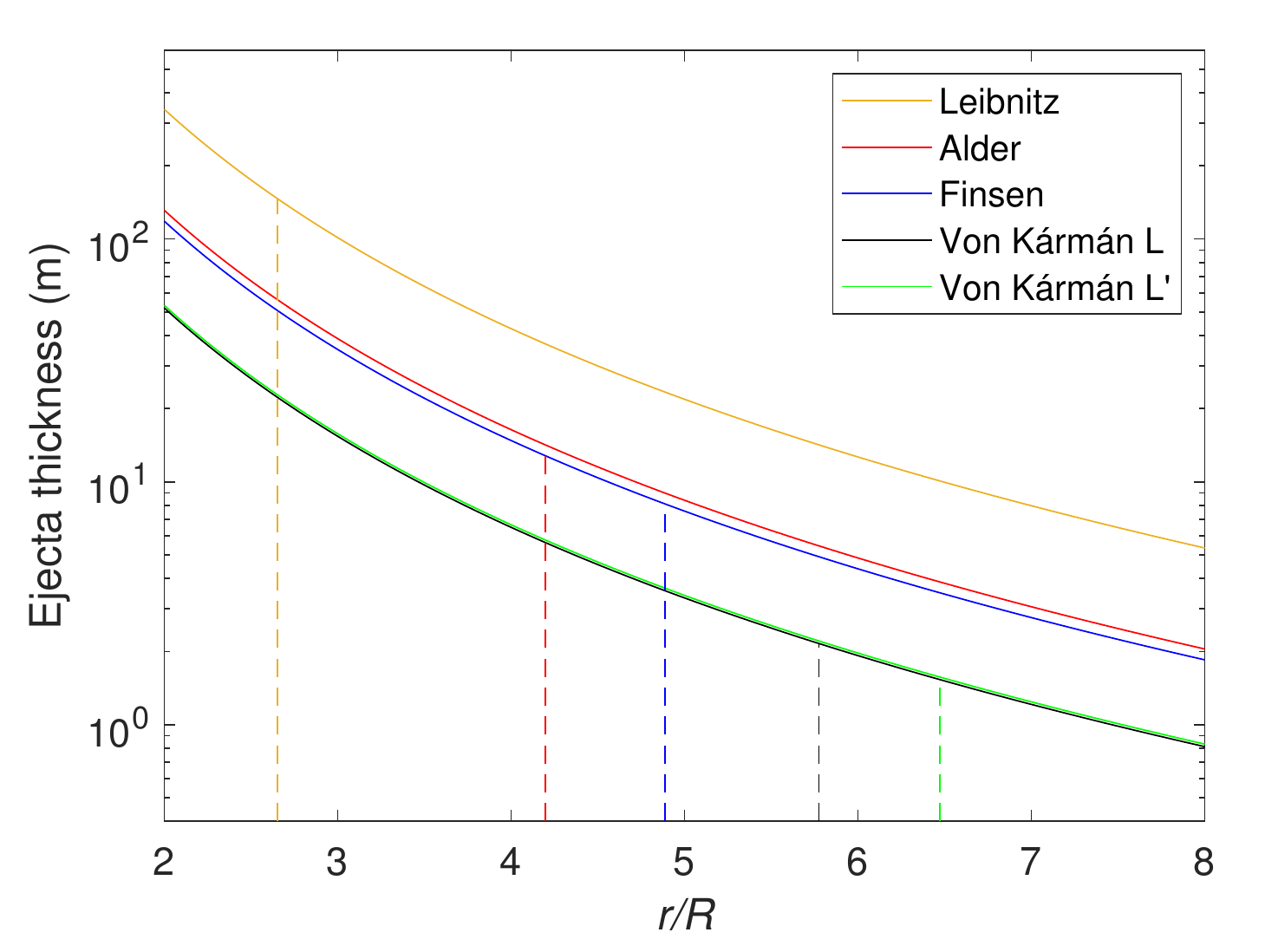}
   \caption{Thickness of distal ejecta from each crater as a function of normalized distance (the ratio of distance to transient radius) from the crater's center. The dashed lines represent the normalized distance at the landing site. The transient radius is calculated according to the model described in \citep{Melosh}.}
              \label{Fig9}%
    \end{figure}

Figure \ref{Fig9} demonstrates the computed ejecta thickness based on this model. The result suggests that $\mathrm{\sim}$ 8.1 m thickness of material comes from Finsen. Finsen is an Eratosthenian crater, and therefore younger than the Imbrian Mare in the Von Kármán crater \citep{Fortezzo,Wilhelms1987}. Figure \ref{Fig1}b indicates the ejecta blanket of the Finsen crater lies on the top of the local regolith and is not overlapped by ejecta from other craters. Therefore, it is reasonable to propose that the majority of top layer L1 is provided by Finsen crater. After billions of years of quiescent soil development and in-situ reworking, the thickness of Layer L1 grows to the current level. Ejecta from Von Kármán L and Von Kármán L' crater are $\mathrm{\sim}$ 1.6 m and $\mathrm{\sim}$ 2.2 m thick, respectively. Since these two craters are of a similar age to Finsen \citep{Fortezzo} and some of their secondary craters are found near the landing region \citep{Hunang}, we believe they contribute the majority of the regolith of two other layers (L2 and L3). According to the ejecta thickness model, the Nectarian Leibnitz crater and Imbrian Alder crater might also transport massive ejected material ($\mathrm{\sim}$ 146 m and $\mathrm{\sim}$ 14 m) at the landing site. However, the latest studies suggest that they are older than the Imbrian mare unit in Von Kármán \citep[e.g.,][]{Fortezzo,Lu}, indicating that material from Leibnitz and Alder may be buried underneath the mare basalt layer \citep{Hunang}. For layers L4--L6, we did not find any potential source from outside the Von Kármán crater and, thus, their source might come from small craters close to the landing site inside this crater. These craters were formed after the emplacement of flood basalt lava flows and then buried by L3, L2, and L1; therefore it is hard to tell their exact locations from the optical images.

\section{Conclusions}

The LPR onboard the Yutu-2 rover of the CE-4 mission provides an effective method to survey the dielectric properties and subsurface structure of lunar regolith on the far side of the Moon. In this study, we estimate the permittivity, evaluate the dielectric loss, and reveal the shallow stratum of regolith at CE-4 landing site using LPR data. 

The average permittivity increases with depth in the upper 4 m of regolith and remains constant in the deeper regolith. The depth-averaged bulk density profile inverted from permittivity is generally consistent with models from Apollo core samples. The dielectric loss is small, with the maximum penetration time of the LPR signal reaching $\mathrm{\sim}$ 40 m depth. The comparison of our results with measurements of Apollo samples shows that the loss tangent of CE-4 regolith is close to that of highland materials. This is also supported by the combination of LPR and CE-2 MRM observations. Both the permittivity and dielectric loss are much lower at the CE-4 landing site than at the CE-3 landing site. This implies that the CE-4 regolith is made up of highly-porous, low-loss, fine-grained materials, while the CE-3 regolith has higher ilmenite content and is dominated by large rock debris ejected from the nearby Ziwei crater. The stratification structure in the post-processed radargram shows local deposit consisting of multiple layers preserves ejecta from several nearby large craters. We suggest that the bowl-shape structure with a d/D ratio of 0.10--0.12 identified in the cross-section is a buried paleo crater. The crater-ejecta thickness model and 415 nm LRO WAC mosaic indicate that the majority of the top layer is ejecta from Finsen crater. The Von Kármán L crater and Von Kármán L' crater also contribute the two additional, thinner layers beneath the top layer.

\begin{acknowledgements}
       The Chang'E-3 and Chang'E-4 Lunar Penetrating Radar data are downloaded from https://moon.bao.ac.cn/web/enmanager/home. The Lunar Reconnaissance Orbiter (LRO) Wide Angle Camera images and the LRO Narrow Angle Camera image are available at http://wms.lroc.asu.edu/lroc.
\end{acknowledgements}

\bibliographystyle{aa} %
\bibliography{Reference} %

\end{document}